\documentclass[preprint,trackchanges]{aastex701}

\newcommand{\cattwo}{Catalog 2 }

\usepackage{amsmath}
\usepackage{amsfonts}
\DeclareMathOperator*{\argmax}{argmax}

\begin{document}
\shorttitle{Estimating Scattering Timescale of FRBs by Transformer}
\shortauthors{B. Kharel et al.}

\title{Multimodal Transformer Based Generic Mixture Density Network for Scattering Timescale Estimation of Fast Radio Bursts
}

\correspondingauthor{Bikash Kharel}
\email{bk0055@mix.wvu.edu}

\author[0009-0008-6166-1095]{Bikash Kharel}
\affiliation{Department of Physics and Astronomy, West Virginia University, PO Box 6315, Morgantown, WV 26506, USA }
\affiliation{Center for Gravitational Waves and Cosmology, West Virginia University, Chestnut Ridge Research Building, Morgantown, WV 26505, USA}
\email{bk0055@mix.wvu.edu}

\author[0000-0001-8384-5049]{Emmanuel Fonseca}
\affiliation{Department of Physics and Astronomy, West Virginia University, PO Box 6315, Morgantown, WV 26506, USA }
\affiliation{Center for Gravitational Waves and Cosmology, West Virginia University, Chestnut Ridge Research Building, Morgantown, WV 26505, USA}
\email{emmanuel.fonseca@mail.wvu.edu}

\author[0000-0003-3821-8112]{Srinjoy Das}
\affiliation{School of Mathematical and Data Sciences,
West Virginia University, PO Box 6310, Morgantown, WV 26506, USA}
\email{srinjoy.das@mail.wvu.edu}

\author[0000-0002-0940-6563]{Mason Ng}
\affiliation{Department of Physics, McGill University, 3600 rue University, Montr\'eal, QC H3A 2T8, Canada}
\affiliation{Trottier Space Institute, McGill University, 3550 rue University, Montr\'eal, QC H3A 2A7, Canada}
\email{mason.ng@mcgill.ca}

\author[0000-0002-7374-7119]{Paul Scholz}
\affiliation{Department of Physics and Astronomy, York University, 4700 Keele Street, Toronto, ON MJ3 1P3, Canada}
\email{pscholz@yorku.ca}

\author[0000-0002-4623-5329]{Mawson W. Sammons}
\affiliation{Laboratoire d'Astrophysique de Marseille, Aix-Marseille Univ., CNRS, CNES, Marseille, France}
\email{mawson.sammons@lam.fr}

\author[0009-0007-5296-4046]{Lordrick Kahinga}
\affiliation{Department of Astronomy and Astrophysics, University of California, Santa Cruz, 1156 High Street, Santa Cruz, CA 95060, USA}
\affiliation{Department of Physics, College of Natural and Mathematical Sciences, University of Dodoma, 1 Benjamin Mkapa Road, 41218 Iyumbu, Dodoma 259, Tanzania}
\email{lkahinga@ucsc.edu}

\author[0009-0004-4176-0062]{Afrokk Khan}
\affiliation{Department of Physics, McGill University, 3600 rue University, Montr\'eal, QC H3A 2T8, Canada}
\affiliation{Trottier Space Institute, McGill University, 3550 rue University, Montr\'eal, QC H3A 2A7, Canada}
\email{afrasiyab.khan@mcgill.ca}

\begin{abstract}
The discovery rate of fast radio bursts (FRBs) continues to increase with the advent of new radio facilities and yet extracting their astrophysical parameters such as scattering timescale ($\tau$) remains a significant bottleneck. Current $\tau$ measurement approaches like fitting analytic template models and scattering aware de-convolution are accurate but slow, sensitive to initialization, limited by low signal to noise and often require manual supervision. These limitations inspired us to explore fast, robust and scalable machine learning methods to estimate the astrophysical parameter value. We present a deep learning approach named Multimodal Transformer Based  Generic Mixture Density Network (MT-GMDN) which ingests FRB dynamic spectrum and its corresponding timeseries profile through parallel transformer encoders, fuses their latent representations and predicts the distribution of $\tau$ with probabilistic output derived from generic mixture-density formulation. This formulation not only estimates the value of $\tau$ but also captures the (zero inflated) nature of FRB populations where a significant fraction of bursts exhibit unresolvable scattering. We trained MT-GMDN on $\sim3500$ FRBs from  CHIME/FRB \cattwo while holding out some fraction of FRBs for validation during training and for testing after the training completes. The model achieves a coefficient of determination ($R^2$) value of $94\%$ on the expected value of $\tau$ for the events with measurable scattering with an excellent recall value of $90\%$ on the test data set. The model was also able to incorporate heteroskedastic errors enabling us the construction of a confidence interval for the predictions. 
\end{abstract}



\section{Introduction} \label{sec:intro}
\begin{figure}
    \centering
    \gridline{
        \fig{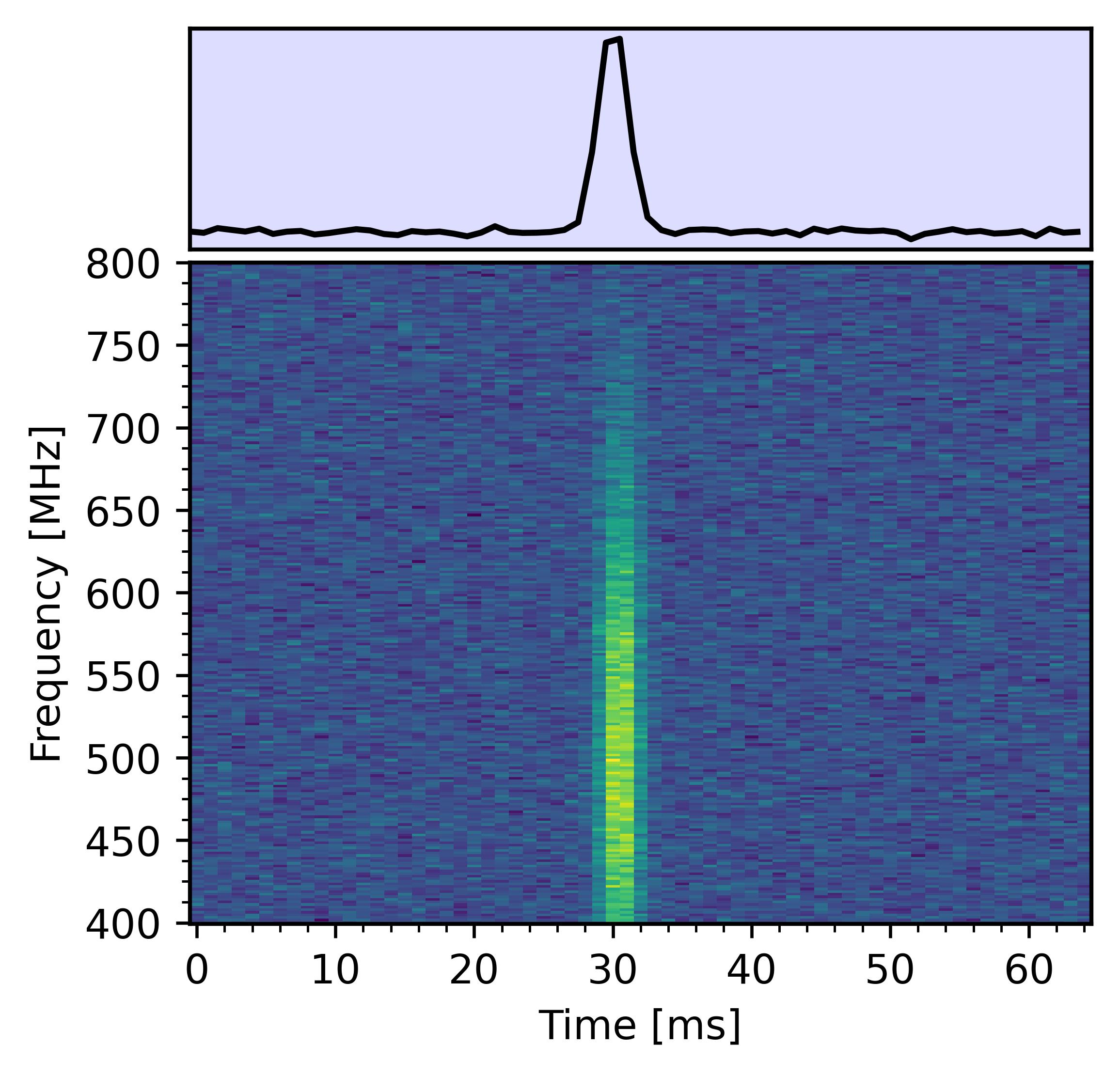}{0.35\textwidth}{}
        \hspace{-5.2cm}
        \fig{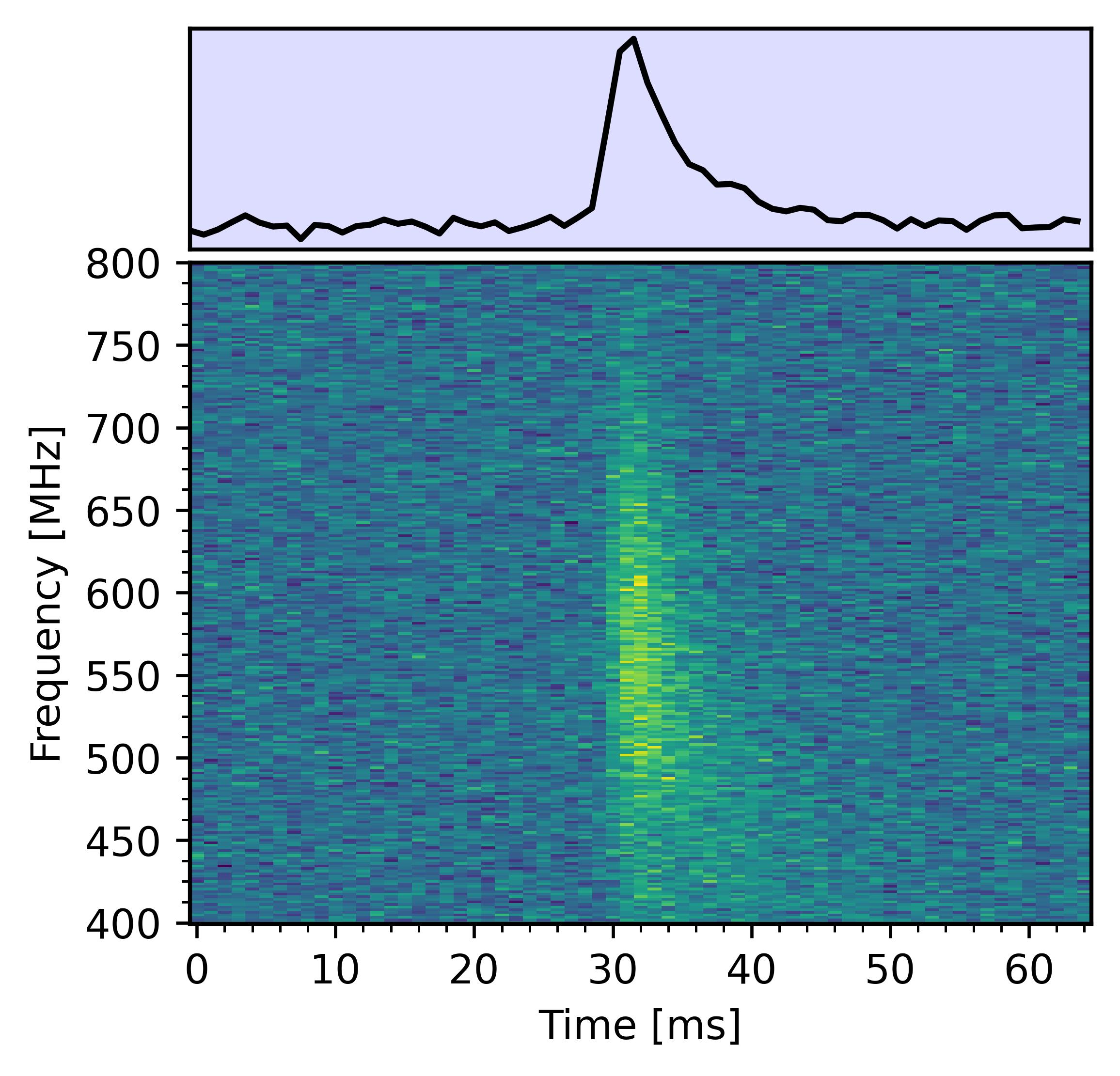}{0.35\textwidth}{}
    }
    \caption{Simulated FRB dynamic spectra with corresponding timeseries at the top of respective dynamic spectrum. The left panel shows an FRB without scattering characterized by a symmetric Gaussian pulse profile, while the right panel shows a scattered FRB with asymmetric pulse profile characterized by exponentially decaying tail. }
    \label{fig:scattering and no scattering}
\end{figure}

Fast radio bursts (FRBs) are powerful extragalactic astronomical transients of millisecond durations \citep{lbm+07} and their observed morphology is strongly determined by propagation effects through ionized interstellar and intergalactic media \citep[e.g.,][]{2003ApJ...596.1142C}. One of the key propagation signatures is a frequency dependent pulse dispersion and temporal broadening due to the multi-path scattering of the radio signal which is often modeled as a one-sided exponential tail with characteristic scattering timescale ($\tau$) \citep{1974MNRAS.166..499W}. Figure \ref{fig:scattering and no scattering} displays two simulated FRBs. The first (left) shows a sharp Gaussian peak with no scattering while the second (right) exhibits an exponentially decaying tail demonstrating scattering.

Electromagnetic radiation from an FRB source traveling through a turbulent interstellar medium (ISM) with inhomogeneous electron density distribution experiences a frequency dependent phase shift. The ``thin screen" model formulates the observed intensity of the scattered radiation
as a result of the phase shift to be an impulse response function (IRF) of the form \citep{2004hpa..book.....L}:
\begin{equation}
\label{eq:decaying exp}
h(t) \propto \exp \left(-\frac{\Delta t}{\tau}\right),
\end{equation}
where,
\begin{itemize}
    \item $\Delta t$ : Time delay between scattered and non-scattered rays and
    \item $\tau$ : Characteristic scattering timescale of the ISM. This scattering timescale exhibits frequency dependence as $\tau \propto f ^ {-4}$
\end{itemize}

The resulting observed pulse profile of an FRB signal is thus the convolution of its intrinsic profile with the scattering exponential kernel, i.e.,
\begin{equation}
    I_{obs}(t) = I_{int}(t) \ast h(t)
\end{equation}

The measurement of $\tau$ carries significant astrophysical implications such as  probing electron density fluctuations along FRB sight lines which thereby constrain the degree of plasma turbulence within host-galaxy environments \citep[e.g.,][]{2016ApJ...832..199X, Shin2025-bp}, diffused halos of intervening galaxies \citep[e.g.,][]{2021ApJ...911..102O} and the large scale ionized structure of the Milky Way \citep[e.g.,][]{2025arXiv250906721S} and constraining redshift \citep[e.g.,][]{Cordes:2021qbo} and the Hubble constant \citep[e.g.,][]{refId0}.

There is a rapid rise in the detections of FRBs with commensal real-time surveys on radio telescopes \citep[e.g.,][]{2018ApJ...863...48C, 2010PASA...27..272M, 2016mks..confE...1J, 2023AAS...24123904S}. The current detection rate is $O(10^3\textrm{ FRBs yr}^{-1})$, which is much larger compared to other astronomical radio transient phenomena. This data volume and diversity of the FRBs in morphology, observational instrument, and propagation effects pose significant challenges for extracting physical parameters such as $\tau$ from the detected FRBs.

Common scattering measurement tools, such as 
\texttt{fitburst} \citep{Fonseca2024}, fit parametric templates to the de-dispersed dynamic spectrum using regression methods. While these techniques can often be accurate, they face significant limitations. Parametric template fitting relies heavily on an adequate initial guess of the parameters and is sensitive to both low signal to noise and complex burst morphologies, thereby often failing to converge. Alternative approaches to estimating scattering timescale include Markov  Chain Monte Carlo (MCMC) methods \citep[e.g.,][]{2019Natur.566..230C} and Fourier Transform based deconvolution methods \citep[e.g.,][]{2003ApJ...584..782B}. However, these methods are computationally slow, require extensive per sample manual intervention and thus are not feasible for analyzing large scale datasets. 

This work mainly relies on the data set from CHIME/FRB \cattwo \citep{2026ApJS..283...34C}, the largest catalog of unique FRBs till date, which includes 4545 FRBs, comprising 3564 one-off events and 981 repeating bursts from 83 different sources. The CHIME/FRB \cattwo data set consists of radiation in the frequency range of 400-800 MHz encoded as dynamic spectrum into 16,384 frequency bins across ~160 time samples of temporal resolution 0.983 ms.

The applications of deep learning in FRB science have been primarily focused on classification \citep[e.g.,][]{Connor_2018, kharel2025repeatingvsnonrepeatingfrbs} and detection \citep[e.g.,][]{Agarwal2020}. As a novel approach in FRB science, to address the issues with methods of scattering timescale estimation discussed in the above paragraph, we present a deep probabilistic model which estimates the distribution of $\tau$ from de-dispersed dynamic spectra and their corresponding pulse profiles. Instead of treating scattering estimation as deterministic regression, we model the conditional distribution $p(\tau | \textbf{x})$, where \textbf{x} is the burst's dynamic spectrum. This distribution approximation yields a well-calibrated heteroskedastic \footnote{Uncertainty parameter is data-dependendent.} uncertainty related to noise associated with the dynamic spectrum containing an FRB signal, allowing one to construct a confidence interval on point estimates.  The deep probabilistic framework not only estimates the distribution of $\tau$, but also quantifies the Bernoulli probability of $\tau\geq0$ as more than $50\%$ of the FRBs have unresolved $\tau$ in CHIME/FRB \cattwo. 

We organize our work in this paper as follows. In Section \ref{sec:methods} we present the methodology underlying our approach which involves probabilistic formulation of scattering timescale estimation, deep learning architecture used to model the formulated distribution and data sources with preprocessing steps applied. Section \ref{sec:training and results} presents training details and results which includes evaluation performed on real observations from CHIME/FRB Catalog 2 through comparison with \texttt{fitburst}, quantitative validation using simulated data, benchmarking our approach against powerful MCMC based inference and investigation of the robustness of our method under different noise conditions. In Section \ref{sec:discussion} we discuss the implications of our findings and in Section \ref{sec:limitation and future work} we outline limitations of our work and directions for future research. Finally, we conclude our work in Section \ref{sec:conclusion}.

\section{Methods} \label{sec:methods}
\subsection{Mixture Density Network Formulation}
Based on the probabilistic interpretation of regression tasks, it is usually assumed that a target t (a variable that a regression methods seek to predict) for given data point $\textbf{x}$ follows a normal distribution
$$
p(t|\textbf{x}) \sim \mathcal{N}(t |\mu(\textbf{x}), \sigma^2).
$$
Where $\mu(\textbf{x})$ is the mean of the target distribution and usually taken as point estimate while $\sigma$ is uncertainty parameter or variance of the target.
Here the uncertainty in the target is homoskedastic\footnote{Variance of the target distribution is data independent}. The regression is then performed by maximizing the likelihood associated with the training dataset. This is written as
$$
\hat{\theta}_{ML} = \argmax_\theta \prod_{i=1}^{N}\mathcal{N}(t_i | y(x_i, \theta), \sigma^2),
$$
where, $y(x_i, \theta)$ is the linear combination of some linear basis functions or neural network with parameters $\theta$. This formulation is overly generic and assumes the distribution of targets to be a normal distribution which may not be always true for a required specific problem domain.

Scattering timescales in FRBs are strictly non-negative (i.e. $\tau \geq 0)$ with the vast majority of FRBs exhibiting minimal scattering with a significantly smaller population displaying large scattering tail. Also, the standard error associated with exponential distributions as in Equation \ref{eq:decaying exp} is proportional to the measured value. Given the wide range of observed $\tau$ for FRBs, it becomes more appropriate to assume the distribution of scattering timescales to follow lognormal distribution instead of normal distribution.
\begin{figure}[ht!]
    \gridline{
        \fig{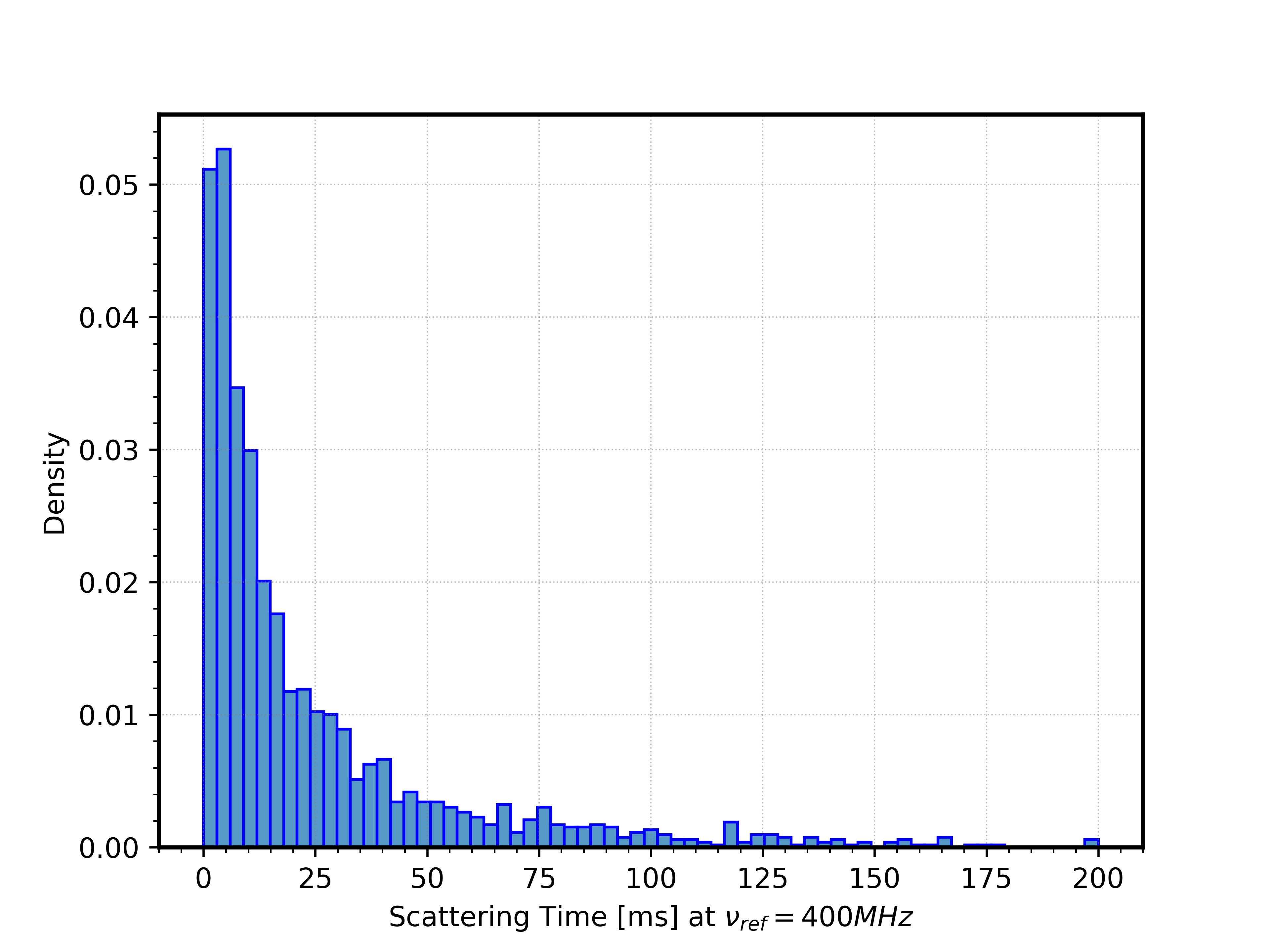}{0.50\textwidth}{}
        \hspace{-1.5cm}
        \fig{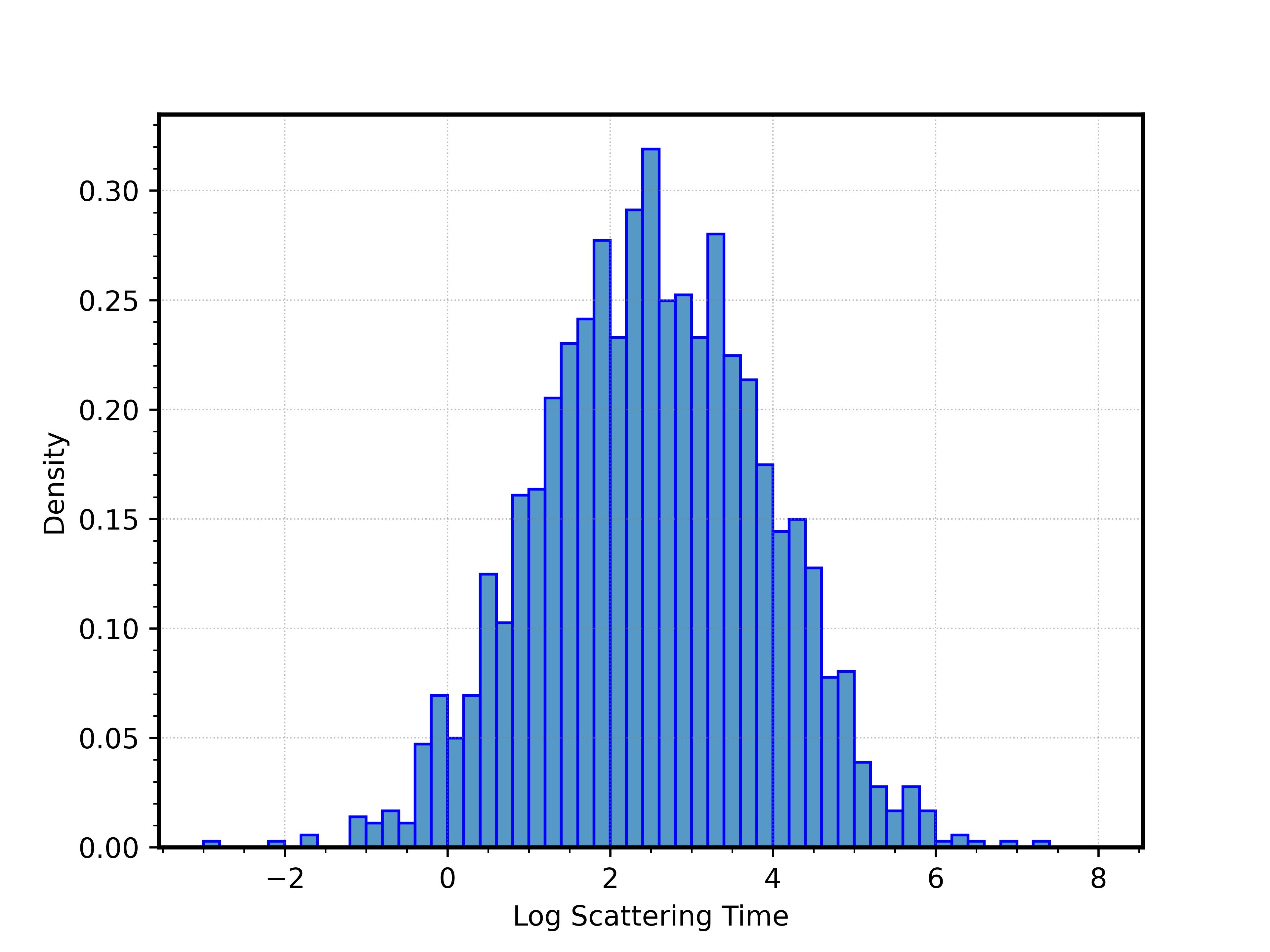}{0.50\textwidth}{}
    }
    \caption{Left: The distribution of $\tau$ from CHIME/FRB \cattwo ( at reference frequency 400 MHz) exhibiting highly skewed heavy tailed profile. Right: Corresponding log transformed ($\ln\tau $) distribution. The exhibition of Gaussian shape in the right panel verify that the scattering timescales follow a lognormal distribution.}
    \label{fig:scattering distribution}
\end{figure}
Figure \ref{fig:scattering distribution} demonstrates that the scattering timescales from CHIME/FRB \cattwo follow a lognormal distribution.
Thus, the distribution of $\tau$ for a given dynamic spectrum \textbf{x} of an FRB event can be expressed as
\begin{equation}
    p(\tau | \textbf{x}) = \textrm{Lognormal}\left(\tau | \mu(\textbf{x}), \sigma^2(\textbf{x})\right) \equiv f_{\textrm{LN}}\left(\tau | \mu(\textbf{x}), \sigma^2(\textbf{x})\right),
\end{equation}
where, $\mu$ and $\sigma^2$ are the location and scale parameters of the lognormal distribution, respectively. Here, the scale of the lognormal distribution is data dependent, i.e., heteroskedastic. This heteroskedastic uncertainty is expected to be larger for dynamic spectrum with low signal to \citep{1997ieas.book.....T} and is also expected to be scaled along with the scattering timescale as standard error of the expected value of exponential tail is directly proportional to the expected value. Therefore, our regression task now becomes maximizing the likelihood associated with the training dataset with the target ($\tau$) distribution being lognormal. This regression task can be written as
\begin{equation}
\label{eq:lognormal likelihood}
    \hat{\theta}_{ML} = \argmax_\theta \prod_{i=1}^{N}f_{\textrm{LN}}\left(\tau_i | \textbf{x}_i; g_{\theta}(\textbf{x}_i)\right),
\end{equation}
where,
$g_{\theta}(\textbf{x}_i)$ can be any arbitrary function which, in our case, is a neural network that approximates $\mu_i$ and $\sigma_i$.

We can now design and train a neural network to maximize the likelihood in Equation \ref{eq:lognormal likelihood}. After training, the network approximates the conditional distribution parameters $\mu_i$ and $\sigma_i$ for any given input $\textbf{x}_i$. However, a notable challenge arises due to CHIME/FRB \cattwo dataset's zero inflation i.e. a large fraction of the population with zero value of scattering timescale. More than $50\%$ of the events in CHIME/FRB \cattwo contains FRBs whose scattering timescales are unresolved (i.e. $\tau \approx 0$), from both repeating. This problem demands standard strategies like discarding the data with unresolved $\tau$ or two staged neural network cascaded such that first predicts the presence of scattering and the second regressing the distribution of $\tau$. Both of the methods are suboptimal as discarding significant fraction of dataset can introduce selection bias and cascading multistage network is computationally redundant and introduces additional complexity. Instead, we propose a generic mixture density network (GMDN) which can be written by modifying a mixture density network \citep{Bishop:DeepLearning24} as 
\begin{equation}
    \label{eq:GMDN}
    p(\tau | \textbf{x}) = \sum_{k=1}^K\pi_k(\textbf{x})f_k(\tau | \textbf{x};g_{k}(\textbf{x})),
\end{equation}

where, $\pi_k(\textbf{x})$ are mixing coefficients and $g_k(\textbf{x})$ are some functions or neural networks. 

In our case, if $p_0(\textbf{x)}$ denotes the probability of $\tau=0$ and $\delta(\tau)$ the probability mass function at $\tau=0$ then
\begin{equation}
\label{eq:mixture components}
\begin{aligned}
& \pi_1(\textbf{x}) = p_0(\textbf{x};g_{\theta}(\textbf{x})) \hspace{0.5cm} \textrm{and} \hspace{0.5cm} f_1(\textbf{x}) = \delta(\tau), \\
& \pi_2(\textbf{x}) = 1 - p_0(\textbf{x}; g_{\theta}(\textbf{x})) \hspace{0.5cm} \textrm{and} \hspace{0.5cm} f_2(\textbf{x}) =  f_{\textrm{LN}}\left(\tau | \textbf{x}; g_{\theta}(\textbf{x})\right).
\end{aligned}
\end{equation}

The neural network $g_{\theta}(\textbf{x)}$ now predicts the Bernoulli probability  of $\tau \geq 0$ as $p_0$ along with  $\mu$ and $\sigma$ parameters for the  distribution of $\tau$ given an input \textbf{x}.

Finally, we can maximize the likelihood of the observed training data under the distribution described by GMDN in Equation \ref{eq:GMDN}, i.e.
\begin{equation}
      \label{eq:maximize GMDN}
      \hat{\theta}_{ML} = \argmax_\theta \prod_{i=1}^{N}\left\{\sum_{k=1}^K\pi_k(\textbf{x}_i)f_k(\tau_i | \textbf{x}_i;g_{k}(\textbf{x}_i))\right\}.
\end{equation}

We have thus formulated a GMDN framework that allows a single neural network to model the presence of scattering in an FRB signal along with the distribution of $\tau$. The estimation of the distribution of $\tau$ inherently accounts for the heteroskedastic uncertainty as well as enabling a robust confidence interval construction. The choice and architecture of the neural network for maximizing the likelihood in Equation \ref{eq:maximize GMDN} is discussed in the following sections.

\begin{figure}[ht!]
    \centering
    \includegraphics[width=0.5\linewidth]{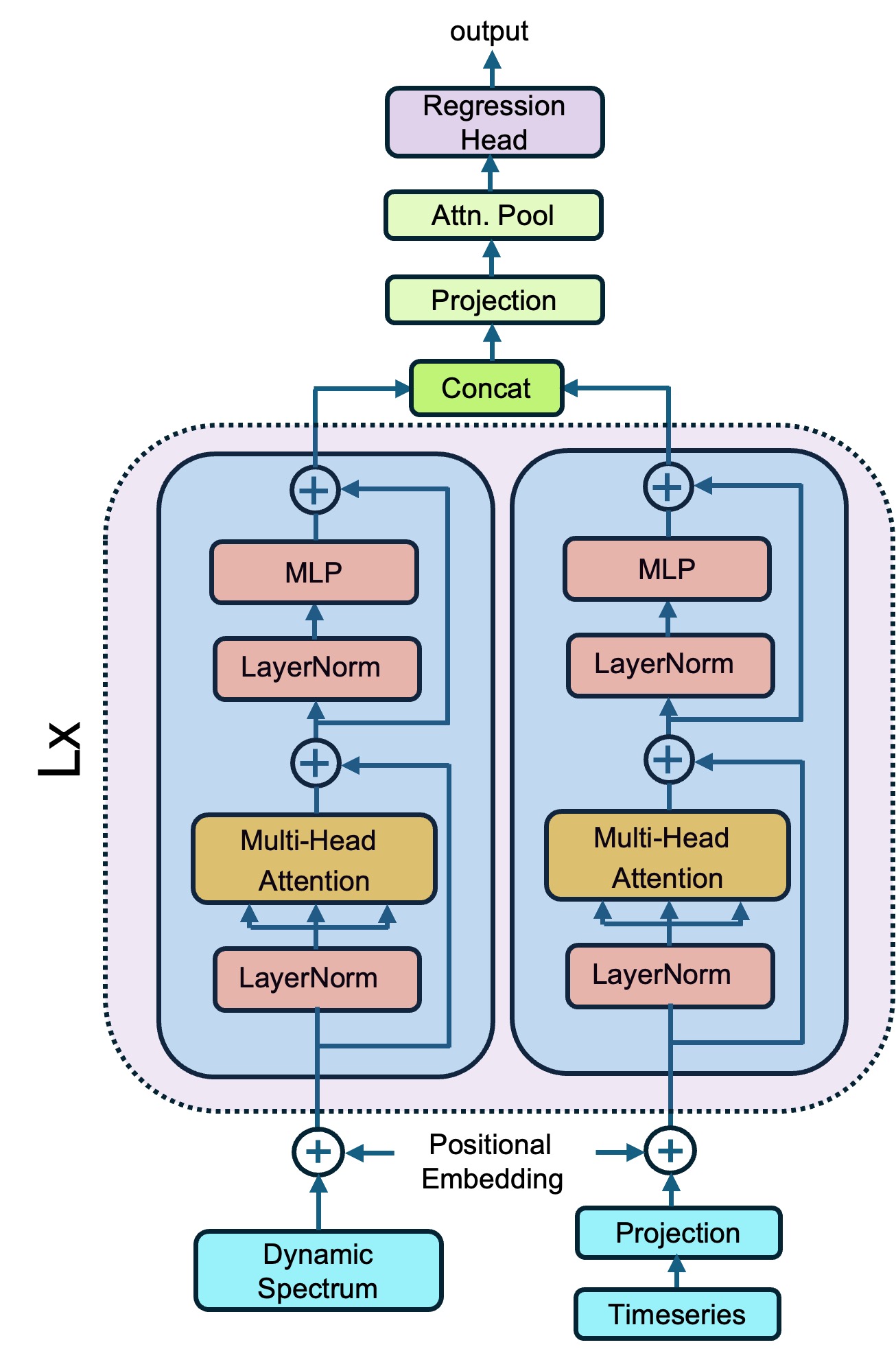}
    \caption{Schematic diagram of the parallel transformer architecture for our regression task. The model processes dynamic spectrum and timeseries representations with two independent parallel transformer branches. The resulting contextual embeddings are then concatenated and passed through a final regression head to predict the output values.}
    \label{fig:parallel transformer}
\end{figure}

\subsection{Model Architecture} \label{sec:model architecture}
We experimented with various neural network architectures, starting with a convolutional neural network (CNN) autoencoder for each dynamic spectrum. CNNs impose translational invariance assuming ``spatial stationarity", referring to uniform distribution of certain features of the image \citep{LeCun2015-wo}, which  is not physically justified for scattering morphology. This is because of the fact that the same scattering morphology at the top of the frequency band corresponds to different scattering timescale than at the bottom of the frequency band. Thus, the CNN based regression failed to give good results with $R^2<50\%$ on the regression task. The vision transformer (ViT) \citep{dosovitskiy2021imageworth16x16words} has been outperforming CNNs in most tasks but it requires very large training dataset to learn the inductive biases associated with the image data.

Due to the limitations with the aforementioned architectures, we designed a transformer architecture for the representation of FRB dynamic spectra where each time bin in dynamic spectrum acts as a token with each token being $d$-dimensional frequency vector as shown in Figure \ref{fig:ds token embedding}. Upon training the transformer, we observed that the model was overfitting by performing exceptionally well on training data while not being able to generalize on test data. We then computed a frequency-averaged timeseries by collapsing the dynamic spectrum along the frequency axis. This operation produced a one-dimensional timeseries with the time samples equal to that of original dynamic spectrum. Each time sample is then projected to a $d$-dimensional vector with a single layer feed forward neural network, thereby creating embedding vectors for each time step. We constructed the same transformer architecture for time series as was developed for the dynamic-spectrum representation described above. However, we observed that the timeseries-based transformer model under-fit the data.

Ultimately, we developed a parallel transformer architecture to process both the dynamic spectrum and time series representations but separately and concatenated the contextual output from both of them. This parallel network achieved a performance superior to all other architectures and methods discussed before and was free from both overfitting and underfitting (see Section \ref{sec:quantitative performance}). The schematic diagram of the multi modal network architecture is shown in Figure \ref{fig:parallel transformer}. A breakdown of the network's constituent components are discussed in Appendix \ref{sec: breakdowm of model architecture}.

\subsection{Data and preprocessing}\label{sec:data and preprocessing}
For this work, we used real data acquired with the CHIME telescope presented in CHIME/FRB Catalog 2. The catalog provides de-dispersed dynamic spectra, $I(\nu, t)$, which are cutouts of the total intensity data. This dynamic spectra capture the intensity of FRB sources at different times ($t$) across different frequency channels ($\nu$). Each dynamic spectrum consists of 16,384 frequency channels spanning the 400-800 MHz band and $\sim$160 time samples with a resolution 0.983 ms. This time window is chosen to fully capture both the burst envelope and any scattering tail with the burst peak lying near the center of the window. We applied 64 channel block averaging to downsample frequency channels to 256, which improved the signal to noise while maintaining computational tractability. Each dynamic spectrum was then independently min-max normalized:
\begin{equation*}
    I'(\nu, t) = \frac{I(\nu, t) - \textrm{min}(I(\nu, t))}{\textrm{max}(I(\nu, t))-\textrm{min}(I(\nu, t))}.
\end{equation*}
The normalization ensures all inputs lie in the range [0, 1]. 

Data for the timeseries transformer can be generated by different dimensionality reduction techniques along the frequency axis as discussed in Appendix \ref{sec: timeseries transformer}. Each timeseries was also independently min-max normalized ensuring all the timeseries points lie in the range [0, 1].

The target $\tau$ values for our regression task are taken from CHIME/FRB \cattwo, which are reported in milliseconds as measured by the modeling tool \texttt{fitburst}. The target values are processed in such a way that $\tau' = \ln(\tau)$ for $\tau > 0$ and $\tau' = 0$ otherwise to compute the objective function in Equation \ref{eq:final neg likelihood}.

We do not apply any explicit filtering of data samples based on signal to noise ratio, burst width or morphology. This broad inclusion of the data enables the model to learn from the full diversity of FRB population present in the CHIME catalog. Besides CHIME/FRB data, we employ synthetic data generated by \texttt{simpulse}\footnote{\href{https://github.com/kmsmith137/simpulse}{simpulse: C++/python library for simulating pulses in radio astronomy}} for the model evaluation purposes. 

\section{Training and Results} \label{sec:training and results}
The MT-GMDN was trained exclusively on data from CHIME/FRB \cattwo. The \cattwo dataset was partitioned into train ($\sim$3800), validation ($\sim$150) and test ($\sim$150) sets for the events with valid scattering timescale values and the events with $NaN$ values of scattering timescale were discarded. Training and validation sets were used while training, and test was used to evaluate the trained model. We randomly sampled the events in the partitions from both repeating and non-repeating sources as the repeating sources also display variability in the scattering \citep[e.g.,][]{2026ApJ..1000L..53P, 2023MNRAS.519..821O}. The complete set of hyperparameter values  adopted for training the MT-GMDN model are detailed in Appendix \ref{sec:hyperparameters}. The final model corresponds to the training step achieving the lowest validation loss which is then used for quantitative evaluation and inferences. 

We used the coefficient of determination ($R^2$) to assess model performance on the regression task, which is defined by
\begin{equation*}
    R^2 = 1 - \frac{SS_{res}}{SS_{tot}},
\end{equation*}
\noindent where,
\begin{itemize}
    \item $SS_{res}=$ Residual sum of squares;
    \item $SS_{tot}=$ Total sum of squares.
\end{itemize}

The classification performance was quantified via a confusion matrix displayed in the Figure \ref{fig:confusion matrix} and the resulting accuracy, recall, precision and F1-scores. 

\subsection{Quantitative performance on the test set from CHIME/FRB \cattwo}
\label{sec:quantitative performance}
For the regression task, we computed $R^2$ between the predicted conditional median scattering timescale value ($e^{\mu}$) and the \cattwo value for the test set. The best value of $R^2$ was obtained to be $94\%$ for the model trained with timeseries computed by spectral averaging discussed in the Section \ref{sec: timeseries transformer} and only the model performance is discussed in this section. The performances of the models corresponding to timeseries by other methods are discussed in Appendix \ref{sec: timeseries transformer}.
\begin{figure}
    \centering
    \includegraphics[width=0.75\linewidth]{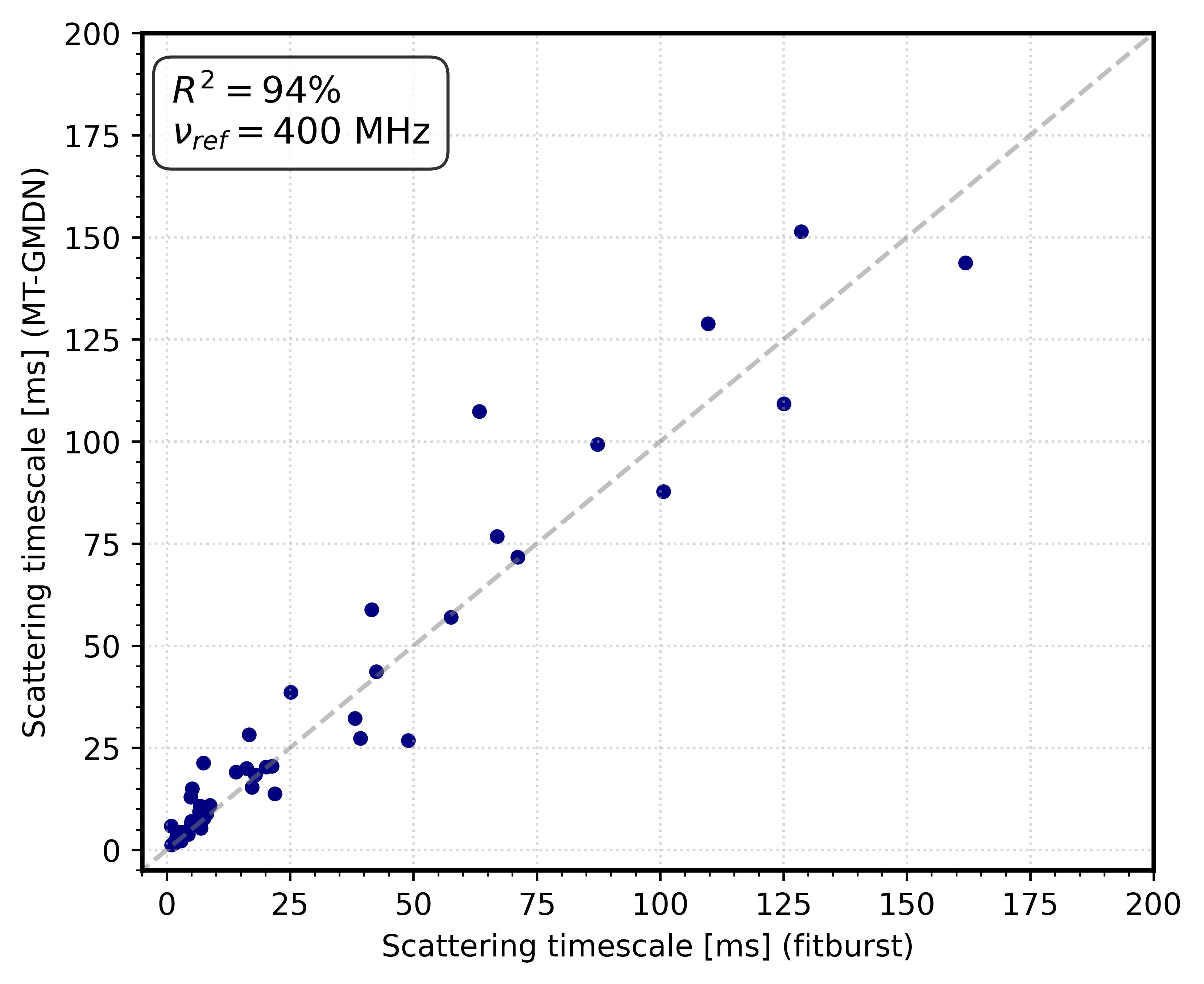}
    \caption{Scatter plot between the MT-GMDN predictions and \texttt{fitburst} measured values of scattering timescales at a reference frequency ($\nu_{ref}$) of 400 MHz. The plot consists of events that have resolved scattering in CHIME/FRB \cattwo.}
    \label{fig:pred vs fitburst}
\end{figure}
    Figure \ref{fig:pred vs fitburst} shows the comparison between the MT-GMDN predicted scattering timescale values against the \texttt{fitburst} measured values in CHIME/FRB \cattwo. The result shows a strong linear correlation with most of the points symmetrically distributed along the $y=x$ line.

Along with the point estimate, we also constructed $95\%$ confidence intervals $e^{\mu \pm 1.96 \times \sigma}$ for the  point estimates.
\begin{figure}
    \centering
    \includegraphics[width=0.75\linewidth]{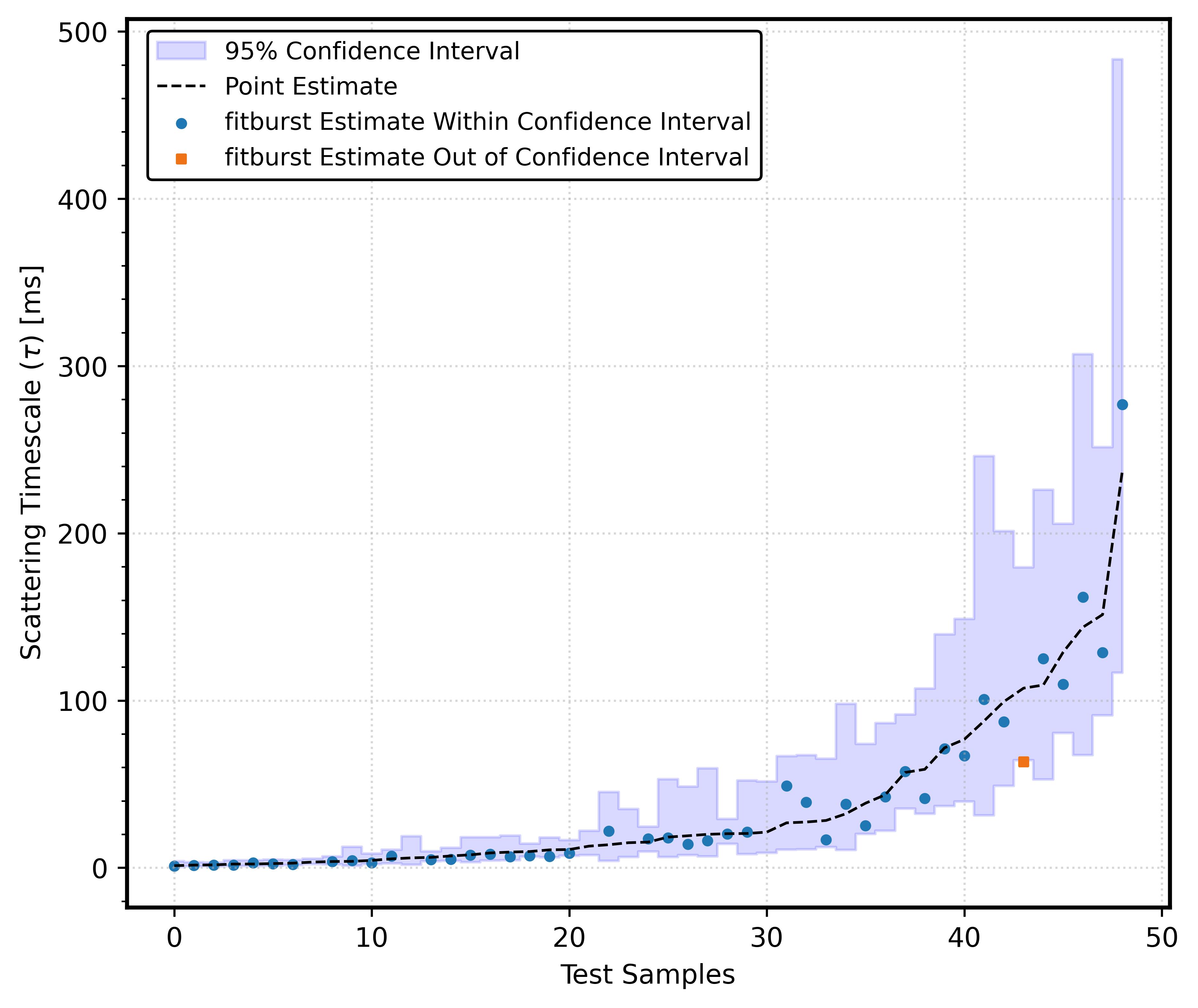}
    \caption{Test set prediction performance with x-axis denoting discrete FRB samples and y-axis scattering timescale at a reference frequency of 400 MHz. \texttt{fitburst} measured values are shown as blue dots and MT-GMDN point estimates are represented by black dashed lines while the blue shaded region representing $95\%$ confidence interval. Orange squares denote the FRB samples which are out of the $95\%$ confidence interval.}
    \label{fig:evaluation_ci_mean}
\end{figure}
Figure \ref{fig:evaluation_ci_mean} presents per sample comparison of MT-GMDN predictions against the \texttt{fitburst} measured cataloged values. The vast majority of samples have \texttt{fitburst} values within the confidence interval of the MT-GMDN point estimate. This indicates the model's reliability not only on the point estimates but also on the well calibrated uncertainty bounds. As illustrated in Figure \ref{fig:evaluation_ci_mean}, the uncertainty associated with the point estimate scales with $\tau$, which is consistent with the characteristic property of the exponential scattering tail. Thus the model was consistently able to recover $\tau$ values along with uncertainty for individual events  across varying  signal characteristics demonstrating its robustness. 

\begin{figure}
    \centering
    \includegraphics[width=0.75\linewidth]{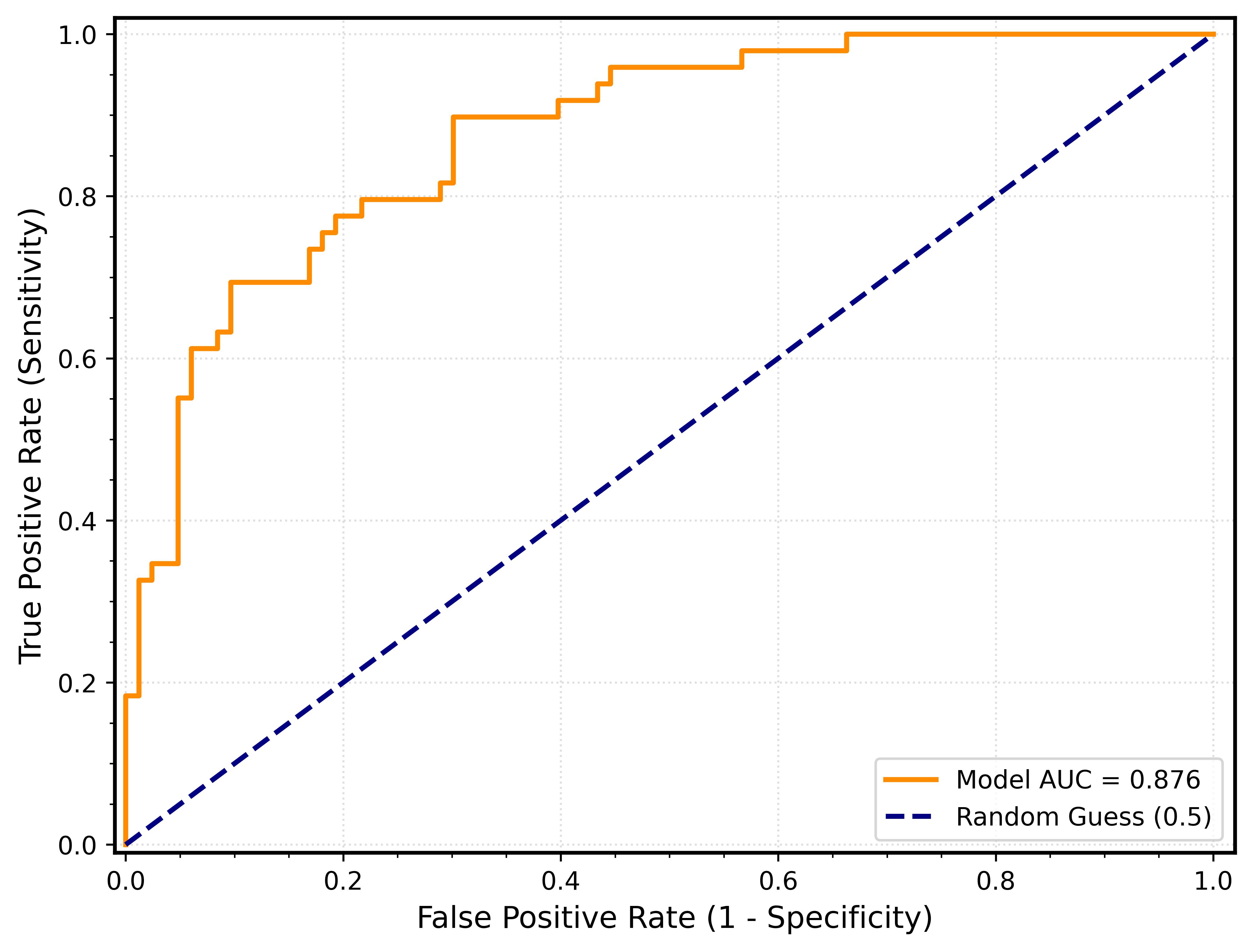}
    \caption{Receiver operating characteristic curve for scattering detection by MT-GMDN at different decision threshold values.}
    \label{fig:roc curve}
\end{figure}

To quantitatively assess the classification performance by the MT-GMDN, we first constructed a receiver operating characteristic (ROC) curve as shown in Figure \ref{fig:roc curve}. This ROC curves yields an area under the curve (AUC) value of 0.87 representing an excellent classifier \citep{c877206e8a3a4ea794f05624f8da6158}. 

\begin{figure}
    \centering
    \includegraphics[width=0.75\linewidth]{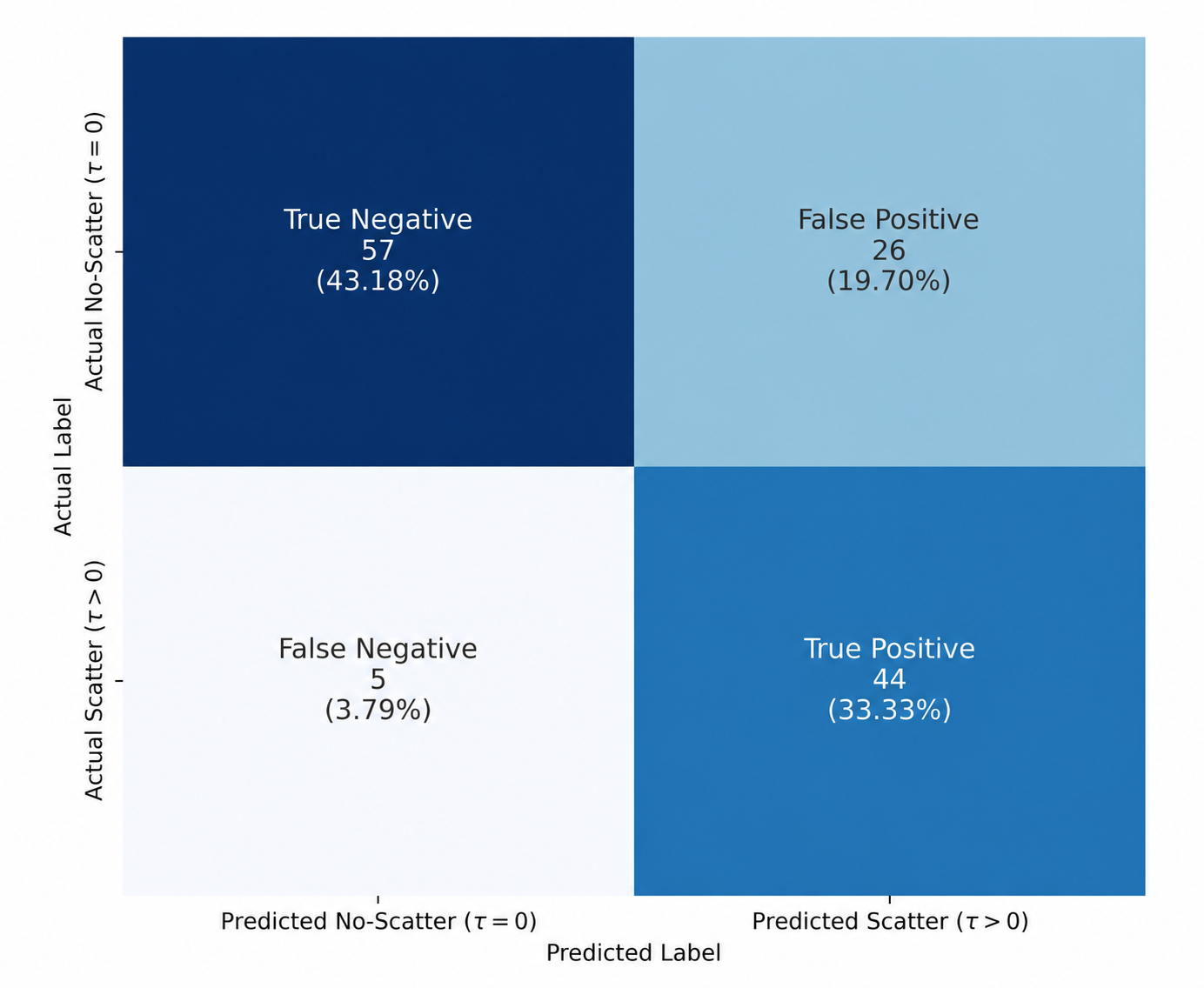}
    \caption{Confusion matrix for the scattering event detection at a detection threshold value of $p_0=0.6$.  Here no-scatter refers to unresolved scattering rather than physical absence of scattering. }
    \label{fig:confusion matrix}.
\end{figure}

Figure \ref{fig:confusion matrix} shows a confusion matrix for the scattering event detection at decision threshold of 0.6 (i.e. if $p_0>=0.6$ then $\tau=0$ else $\tau>0$).
At our chosen threshold, recall value of $90\%$ was achieved and the overall accuracy was $76\%$. The precision of $63\%$ suggest that the model is sensitive to scattering, and it predicted scattering in profiles that were unresolved in CHIME/FRB \cattwo. This conservative approach is critical for ensuring that faint scattering signatures are not overlooked. The decision threshold is an adjustable parameter that can be tuned based on the specific objective of the analysis. For the most balanced performance, we can utilize the ROC curve to determine the optimal value of decision threshold. 

\subsection{Quantitative performance on simulated data}
\label{sec:quantitave performance on simulated data}
To analyze the robustness of the MT-GMDN predictions, we evaluated the performance of the model against a synthetic dataset generated with \texttt{simpulse} framework. We observed a consistent systematic bias in the recovered $\tau$ values, where the model was overestimating the predictions relative to the injected ground truth. The systematic bias was observed when modeling with \texttt{fitburst}. This discrepancy likely originates from overestimation of the scattering effect within the \texttt{simpulse} simulation backend and/or a fundamental morphological mismatch between the templates used by \texttt{fitburst} from which our training target labels were derived and numerical methods implemented in the \texttt{simpulse} backend. 

We accounted this systematic offset by a linear calibration of the model's point estimate and upper and lower bound of the confidence interval. The linear calibration was obtained by modeling zero-noise dynamic spectra with \texttt{fitburst} for the minimum and maximum values of scattering timescales in the simulated data and then calculating the mean offset factor between the measured and injected values.
\begin{figure}
    \centering
    \includegraphics[width=0.75\linewidth]{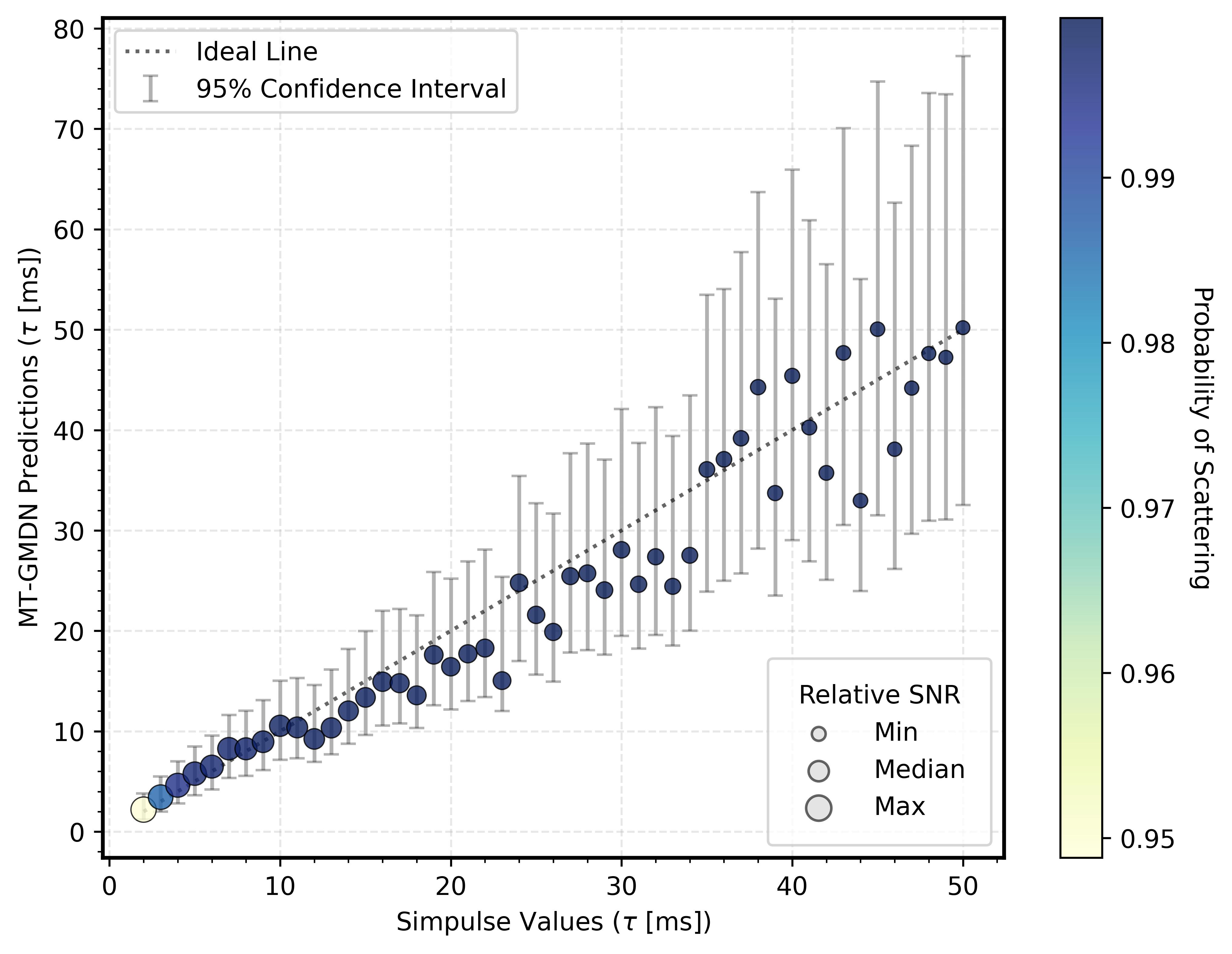}
    \caption{Scattering timescale recovery for synthetic broadband pulses generated by \texttt{simpulse} with intrinsic widths of 1 ms. The vertical error bars denote $95\%$ confidence interval with all of the target values within the interval. The MT-GMDN model achieves $R^2$ score of $92\%$ on the point estimate of the $\tau$ values. Measurement uncertainty scales with $\tau$  justifying the inherent difficulty in characterizing the morphology of the highly scattered profiles.}
    \label{fig:model performance on synthetic data}
\end{figure}
As illustrated in Figure \ref{fig:model performance on synthetic data}, the systematic offset correction aligns the predictions with the ideal line ($y=x$), along with the injected target values falling entirely within the predicted $95\%$ confidence interval. Furthermore, the MT-GMDN captures the measurement uncertainty which are represented by the error bars. These uncertainties scale proportionally with the $\tau$ values and is physically expected as higher $\tau$ values dominate the pulse morphology and suppresses the signal to noise of the burst. The successful capture of the target values with heteroskedastic uncertainty confirms that the MT-GMDN provides both accurate point estimate and reliable confidence interval for scattering timescale.

\subsection{MT-GMDN vs. \texttt{fitburst} on simulated data}
\label{sec:mt-gmdn vs fitburst}
The process of model fitting with \texttt{fitburst} involves two steps. At first, a model without scattering is fitted and the parameters obtained from the first fit is then used to create a model with scattering. Chi-square ($\chi^2$) statistic for each of the fit is computed. These $\chi^2$ statistics are then used to obtain F-statistic
\begin{equation*}
F = \frac{\chi_1^2 - \chi_2^2}{\chi_2^2} \frac{N_{\rm dof, 2}}{\Delta N_{\rm fit}},
\label{eq:f_test}
\end{equation*}
where $N_{\rm dof, 2}$ is the number of degrees of freedom in the complex model with scattering and $\Delta N_{\rm fit}$ is the difference in the number of free parameters between the two models. Using the F-statistic, $p$-value is obtained which represents the improvement in the fit quality after the introduction of the scatter broadening parameter in the model. 
\begin{figure}
      \gridline{
        \includegraphics[width=0.47\textwidth]{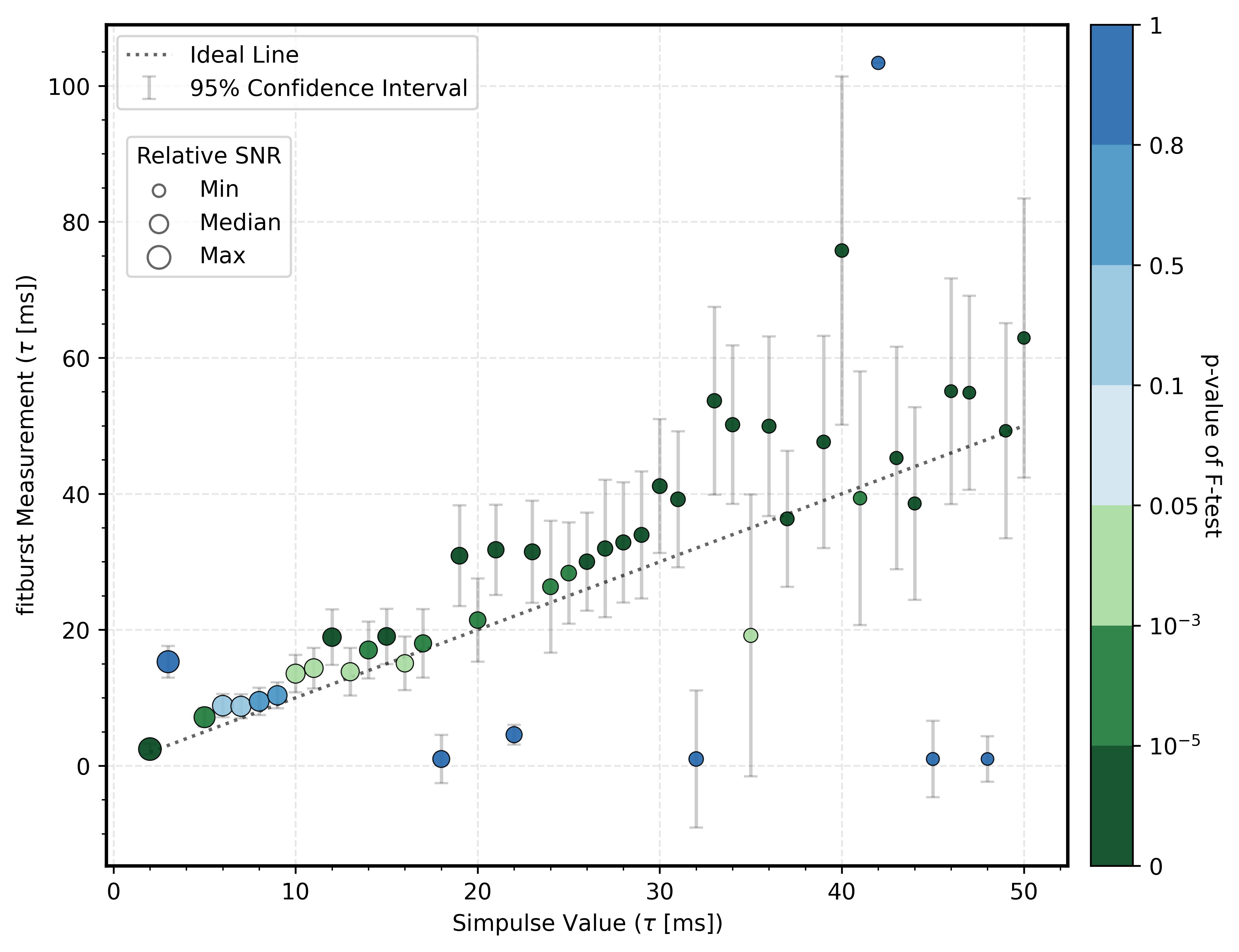}
        \hspace{-1.2cm}
        \includegraphics[width=0.47\textwidth]{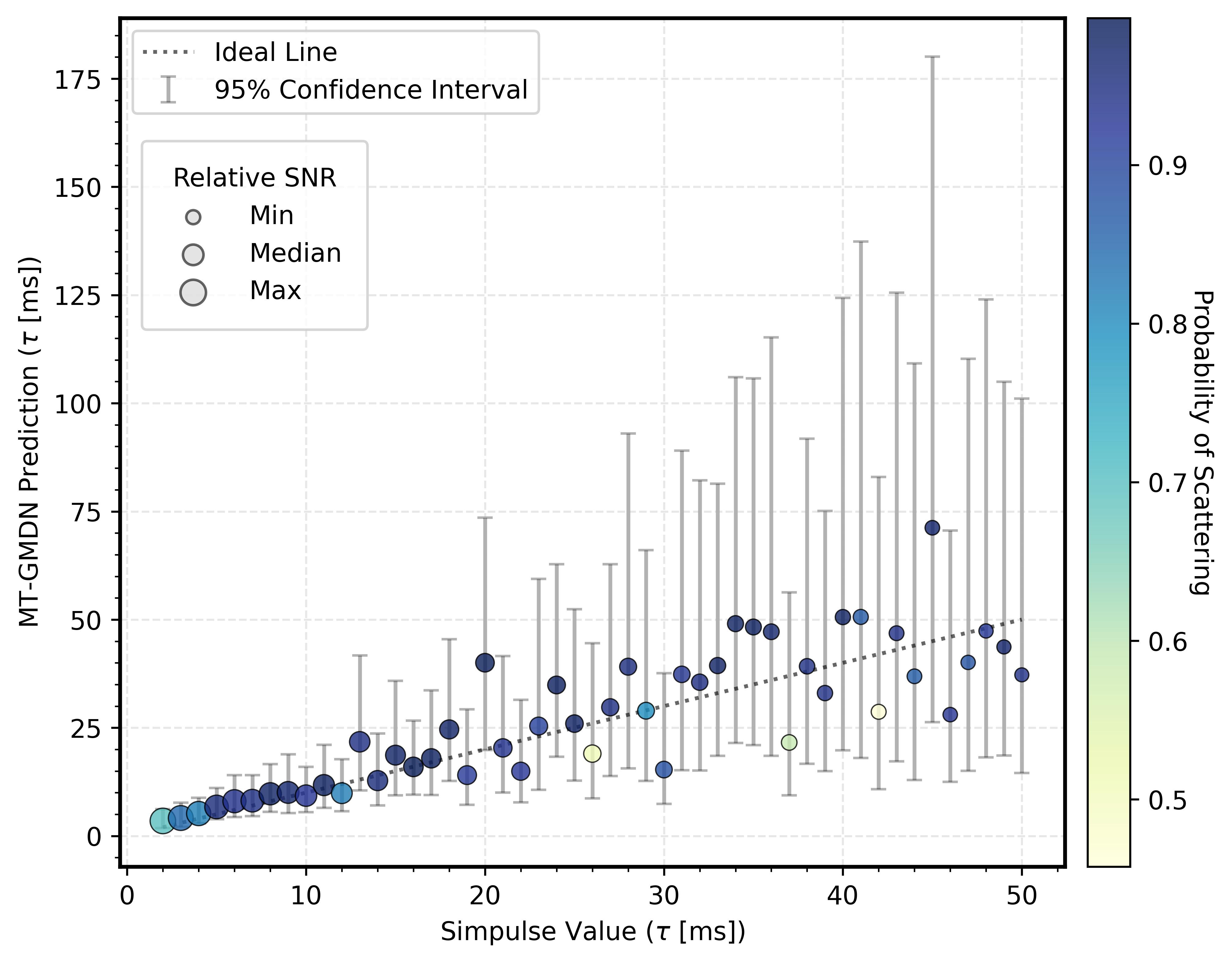}
        }
    \caption{Comparison of \texttt{fitburst} measured values of $\tau$ on synthetic data against MT-GMDN predictions. The synthetic data set consists of broadband pulses with maximum SNR for the event with minimum scattering. Fluence for each of the event is held constant such that the SNR gets lower for the events with higher scattering due to the redistribution of radiation energy across time in exponentially decaying tail. Left: \texttt{fitburst} measurement with the vertical bars representing $95\%$ confidence interval, the size of the markers corresponding to relative SNR and the colorbar represents $p$-value of F-test for scattering vs non-scattering model fit. Right: MT-GMDN predictions with vertical bars capturing $95\%$ confidence interval with the marker size corresponding to relative SNR while the marker color representing the probability of scattering. A linear correction factor was applied to both \texttt{fitburst} and MT-GMDN methods. The maximum signal to noise value of the synthetic data was 20 corresponding to the left most point in the plots and minimum was 4 corresponding to the right most point in the plots. }
    \label{fig:fitburst vs mtgmdn}
\end{figure}

To compare the robustness of our MT-GMDN against \texttt{fitburst}, we created common sets of broadband synthetic events and applied both \texttt{fitburst} and MT-GMDN methods to recover the $\tau$ values. \texttt{fitburst} performed exceptionally well on the data sets where the signal to noise of the events were high and so did MT-GMDN. However, in the low SNR data set where the scattering tails often get mixed up with the background noise, MT-GMDN outperforms \texttt{fitburst} which is illustrated in Figure \ref{fig:fitburst vs mtgmdn}. To quantify the performance difference between \texttt{fitburst} and MT-GMDN we again employed $R^2$ score as a metric for comparison. $R^2$ score of $39\%$ was obtained for \texttt{fitburst} measured values while the $R^2$ score was $68\%$ for MT-GMDN predicted values. Moreover, Figure \ref{fig:fitburst vs mtgmdn} (Left) for \texttt{fitburst} measurement shows significant number of events for which the $p$-value of F-test are significantly high thereby failing to reject the null hypothesis that the non-scattering model is better fit for data. For some of the events, target values fall out of the $95\%$ confidence interval and uncertainties do not scale well with noise levels and $\tau$ values. Figure \ref{fig:fitburst vs mtgmdn} (Right) shows that all the target values fall within the confidence interval predicted by MT-GMDN with uncertainties scaling well with noise levels and $\tau$ values. The $95\%$ confidence intervals are very broad for MT-GMDN method as compared to \texttt{fitburst} at very low signal to noise values  of the synthetic bursts ($<$8 for bursts with \texttt{simpulse} injected value of $>15$ ms), however the point estimates and lower bounds are robust as compared to \texttt{fitburst}.

Besides the performance on recovering $\tau$ values from data, MT-GMDN also has computational advantage. \texttt{fitburst} relies on non-linear least squares optimization for each event which is an iterative process and the algorithm time complexity scales up with the number of frequency channels ($N_{\nu}$), time samples ($N_t$) and number of iterations to converge. After the convergence, \texttt{fitburst} computes exact Hessian and inverts it to obtain the covariance matrix and hence uncertainty. This operation also scales with $N_{\nu}$ and $N_{t}$. However, once trained, MT-GMDN has the time complexity of  $\mathcal{O}(1)$ regardless of the complexity of a burst. It has been observed that MT-GMDN achieves inference speed of approximately four orders of magnitude ($10^4$) faster than \texttt{fitburst} on the retrieval of $\tau$ value and the corresponding uncertainty, enabling real time parameter estimation and uncertainty quantification.

\subsection{MT-GMDN vs. MCMC on simulated data}
MCMC can be used to sample from the posterior distribution in Bayesian approach of parameter estimation when marginal likelihood (evidence) over all the possible values of parameters in high dimensional parameter space becomes intractable. We used the same mathematical model as used by \texttt{fitburst} to model the scatter broadening in pulse profile \citep{McKinnon_2014} given as
\begin{equation}
    T_{kn} = \left( \frac{\nu_k}{\nu_r} \right)^{-\delta} \exp \left[ \frac{\sigma^2}{2\tau_k^2} - \frac{(t_{kn} - t_{0})}{\tau_k} \right] \left[ 1 + \text{erf} \left( \frac{t_{kn} - (t_{0} + \sigma^2 / \tau_k)}{\sigma \sqrt{2}} \right) \right].
    \label{eq:mckinon model}
\end{equation}
Here, 
$T_{kn}$ is the pulse profile value at $k^{th}$ channel and $n^{th}$ time bin.  $\sigma$ is the Gaussian width of the pulse, $t_0$ is arrival time of the pulse, $\tau_k$ is scattering timescale and $\delta$ is the index of scattering accounting for frequency dependence of scattering and was set to  4 which is the empirical standard value for FRBs. $\nu_r$  is the reference frequency which was fixed to 400 MHz for our analysis.

The spectrum was modeled by using the running power law model given as 
\begin{equation}
F_k = \left( \frac{\nu_k}{\nu_r} \right)^{\gamma + \beta \ln \left( \frac{\nu_k}{\nu_r} \right)}.
\label{eq:spectrum}
\end{equation}
Here, $\gamma$ is spectral index and $\beta$ is spectral running. 

Finally, a two dimensional dynamic spectrum $D_{kn}$ was then modeled by combining Equations \ref{eq:mckinon model} and \ref{eq:spectrum} as 
\begin{equation}
    M_{kn} = AF_kT_{kn},
    \label{eq:dynamic spectrum model}
\end{equation}
with $A = 10^{\alpha}$, the amplitude of overall wave packet characterized by global amplitude parameter $\alpha$. 

\begin{figure}
    \centering
    \includegraphics[width=0.85\linewidth]{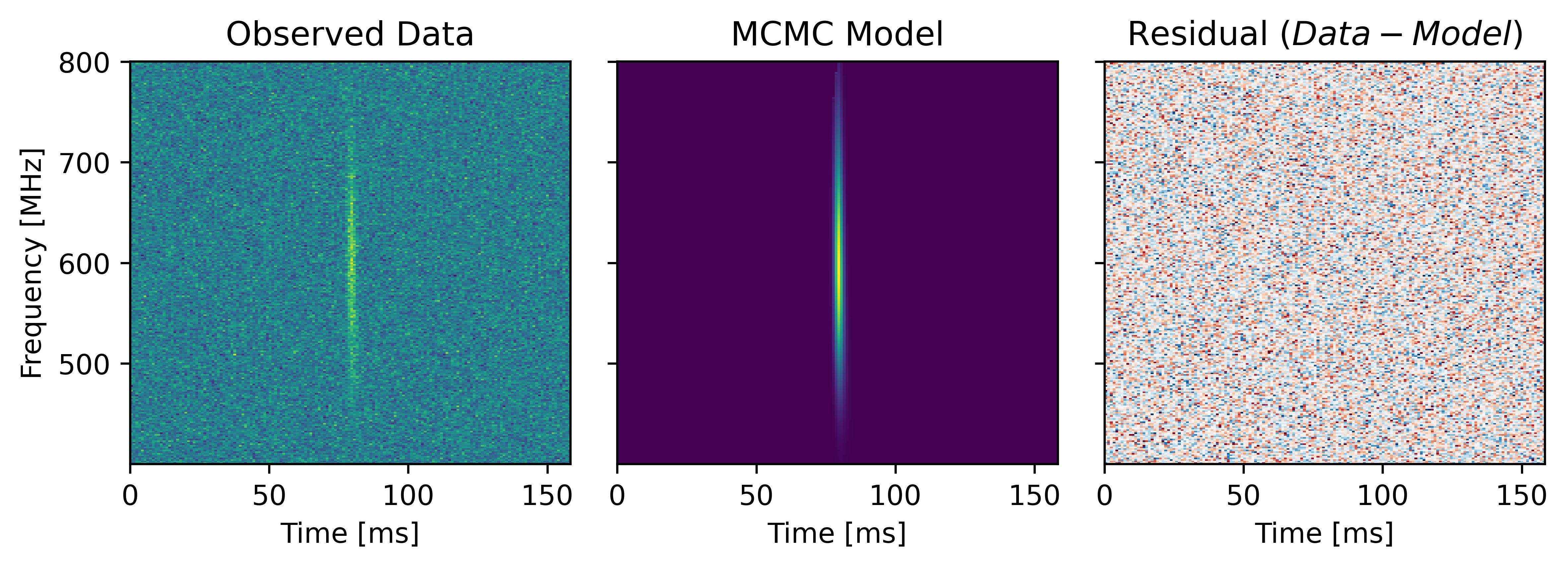}
    \caption{Dynamic spectrum of a synthetic FRB with modeled by MCMC method. The injected pulse had an intrinsic width of 1 ms and scattering timescale of 2 ms.}
    \label{fig:mcmc model residual}
\end{figure}

\begin{figure}
    \centering
    \includegraphics[width=0.85\linewidth]{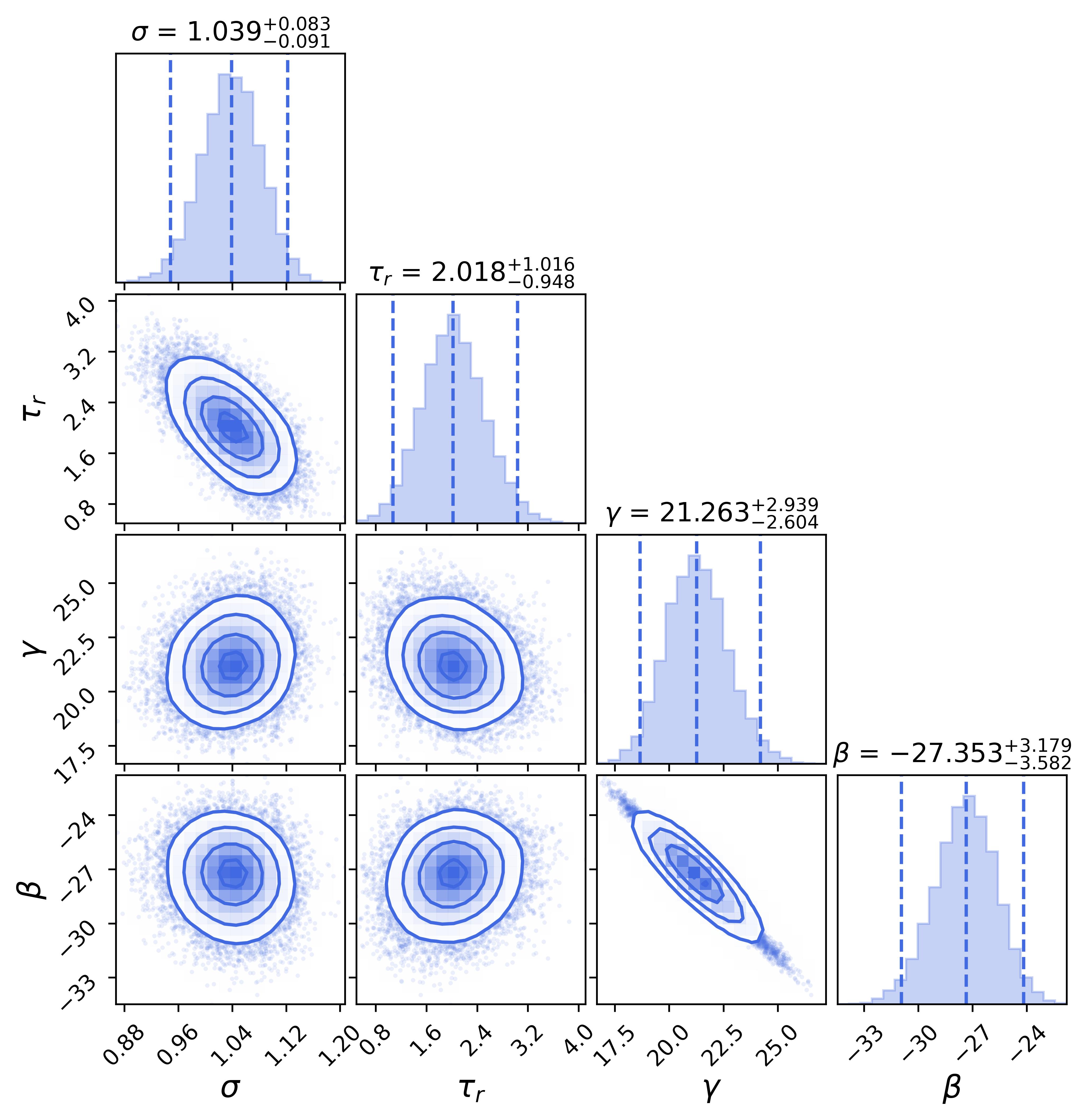}
    \caption{Posterior distribution of some of the model parameters in Equation \ref{eq:dynamic spectrum model} generated by  MCMC method for a simulated FRB shown in the Figure \ref{fig:mcmc model residual}. The central vertical line in the corner plots represent median or the point estimate while the first and third vertical lines denote the $95\%$ credible interval. The posterior distributions are obtained by combining the samples from all MCMC sampler chains. The simulated burst had width of 1 ms and scattering time of 2 ms and MCMC method successfully recovered those values as seen in the corner plots for $\tau_r$ as scattering time and $\sigma$ as the width.}
    \label{fig:mcmc parameter distributions}
\end{figure}

To evaluate the MT-GMDN performance against MCMC on estimating the scattering timescale, we needed to create complete model of each dynamic spectrum in this evaluation. An example of a model generated by MCMC method is shown in Figure \ref{fig:mcmc model residual}. Figure \ref{fig:mcmc parameter distributions} illustrates the parameter distributions along with their corresponding $95\%$ credible intervals for the simulated FRB previously displayed in Figure \ref{fig:mcmc model residual}.

\begin{figure}
    \centering
    \includegraphics[width=0.75\linewidth]{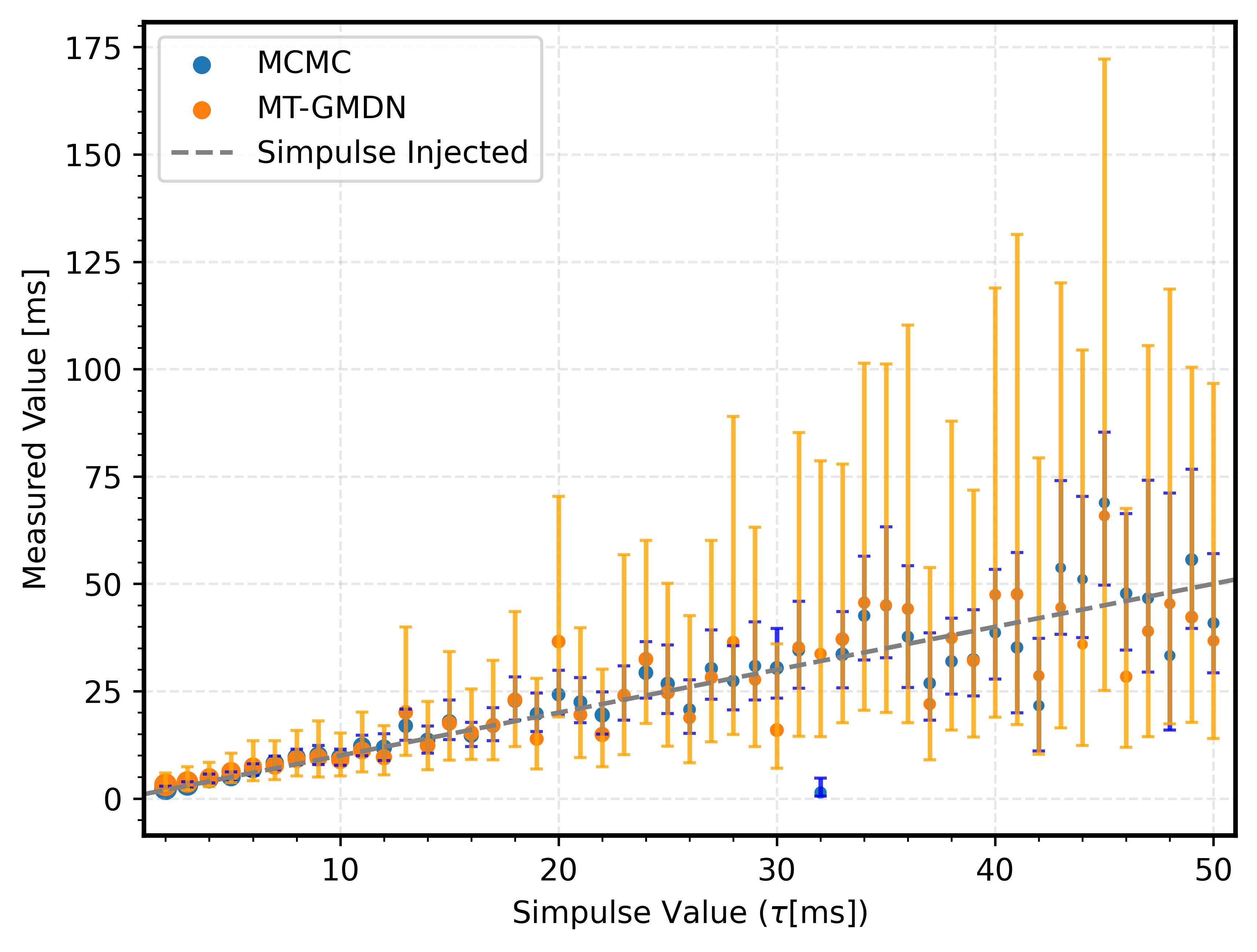}
    \caption{Comparison of MCMC vs MT-GMDN performance on the simulated dataset. The size of the markers denote the SNR of corresponding dynamic spectrum and it ranges from 20 to 4 with larger size denoting higher SNR. The error bars for MT-GMDN predictions represent $95\%$ confidence interval and the error bars for MCMC estimates represent $95\%$ credible interval. The pulse width of all the bursts were set to 1 ms. Both the MCMC and MT-GMDN estimations were systematic bias corrected as introduced by \texttt{simpulse}.}
    \label{fig:mcmc vs mtgmdn}
\end{figure}

We modeled ~50 simulated FRBs with MCMC using lognormal distribution as prior for the $\tau$ and extracted the posterior distribution for $\tau$ by combining all MCMC chains. The heteroskedasticity was incorporated in the MCMC modeling by passing the noise value of a dynamic spectrum directly to the Gaussian likelihood of the MCMC. The sampler used was ``No-U-Turn Sampler"  \citep{2011arXiv1111.4246H} with a burn-in period of 2000 and 4 parallel chains. The comparison plot for MCMC method and MT-GMDN methods is shown in Figure \ref{fig:mcmc vs mtgmdn}.

Some of the key metrics comparing the performance of MCMC versus MT-GMDN on both point estimate and uncertainty are summarized in the Table \ref{tab:mcmc vs mtgmdn}. MT-GMDN performs slightly better on root mean square error (RMSE) and $R^2$ score indicating improved predictive accuracy and better explained variance than MCMC. Both of the approaches also demonstrate negligible bias in the point estimation with MT-GMDN slightly better than MCMC. The significant difference between  MT-GMDN and MCMC methods is on the the uncertainty modeling behaviors.  MT-GMDN exhibits over-coverage, potentially due to conservative estimation of variance in the mixture density framework. However, under-coverage in uncertainty by the MCMC method indicates narrow predictive intervals reflecting its limitations in capturing the full variability of the data. Despite MCMC producing narrower uncertainty intervals, its under-coverage indicates that these intervals are not reliable while MT-GMDN prioritize more conservative but statistically consistent uncertainty estimates at the expense of sharpness. MT-GMDN thus demonstrates competitive or superior point-estimates and lower bounds of the uncertainty interval. 
\renewcommand{\arraystretch}{1.5} 
\begin{deluxetable*}{llclr}
\tablecaption{MT-GMDN vs. MCMC Comparison Metrics \label{tab:mcmc vs mtgmdn}}
\tablewidth{0pt}
\tablehead{
\colhead{Metric} & \colhead{Equation} & \colhead{MT-GMDN} & \colhead{MCMC} & \colhead{Comments}
}
\startdata
RMSE  &  $\left[\frac{1}{N}\sum_{i=1}^{N}(\hat{y}_i-y_i)^2\right]^{1/2}$ & 7.62 & 7.83 & Lower is better \\
$R^2$ & $1-\frac{\sum_{i=1}^{N}(y_i-\hat{y}_i)^2}{\sum_{i=1}^{N}(y_i-\bar{y})^2}$ & 0.71 & 0.69 & Higher is better \\
Bias & $\frac{1}{N}\sum_{i=1}^{N}(\hat{y}_i-y_i)$ & -0.003 & -0.068 & Ideal $\approx 0$\\
$95\%$ Coverage & $\frac{1}{N}\sum_{i=1}^{N}\mathbb{I}(y_i \in [l_i,u_i])$ & 1 & 0.89 & Ideal $\approx 0.95$\\
MIW & $\frac{1}{N}\sum_{i=1}^{N}(u_i-l_i)$ &  19.69 & 2.49 & Lower is better if calibrated
\enddata

\tablecomments{RMSE: Root mean square error, MIW: Mean interval width, $\hat{y}_i$: Point estimate, $y_i$: Target, $u_i$: Upper bound estimate, $l_i$: Lower bound estimate }
\label{tab: mcmc vs mtgmdn}
\end{deluxetable*}

\subsection{Heteroskedastic uncertainty modeling by MT-GMDN}
To assess the uncertainty estimation nature of MT-GMDN based on the amount of noise in a dynamic spectrum, we analyze the correlation between the predicted uncertainty parameter ($\sigma$) and SNR of some simulated bursts. Higher SNR samples are expected to yield more precise estimates while lower SNR events are expected to produce broader uncertainty interval. This implies a negative correlation between predictive uncertainty and SNR of FRB events.

\begin{figure}
    \centering
    \gridline{
        \fig{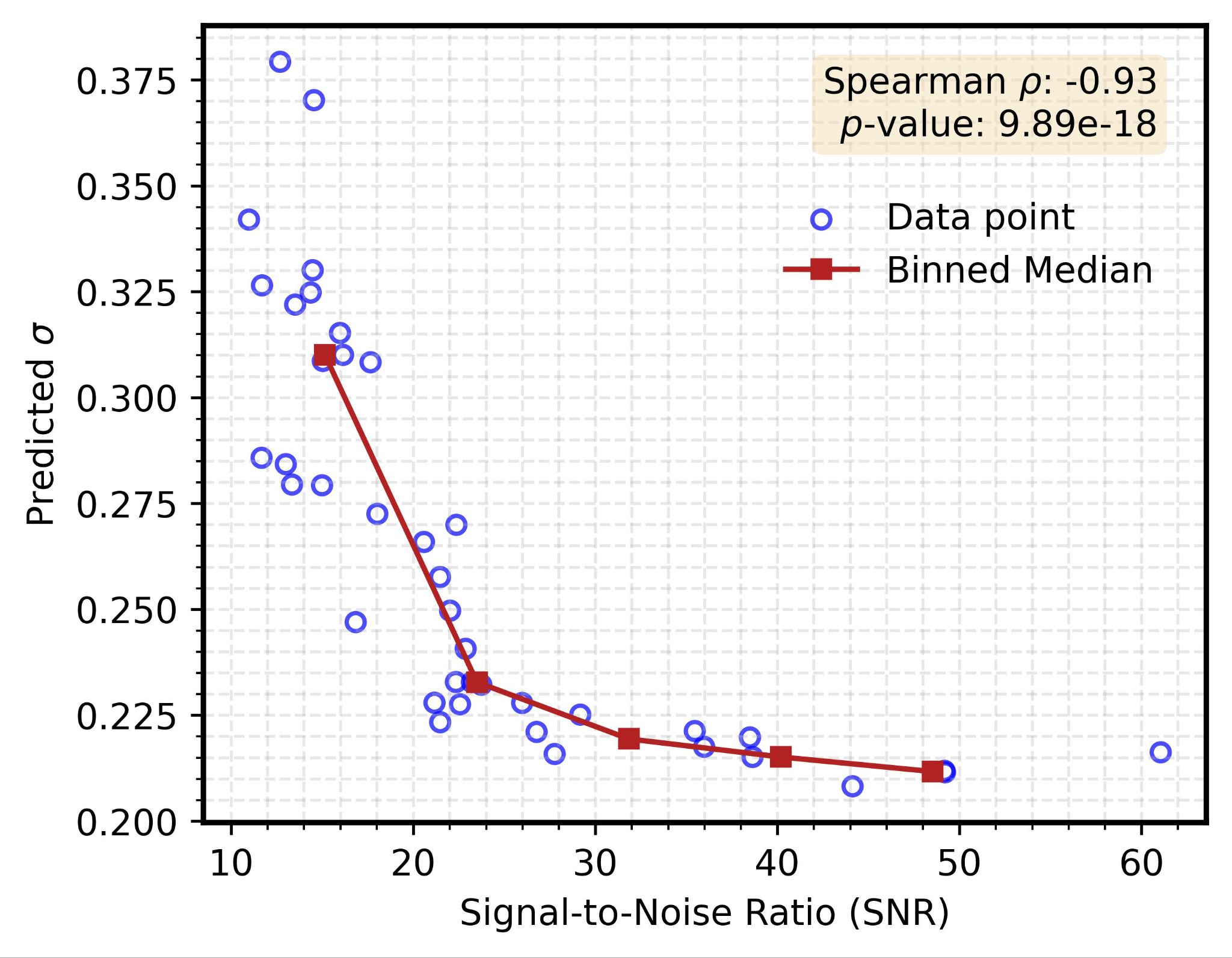}{0.47\textwidth}{}
        \fig{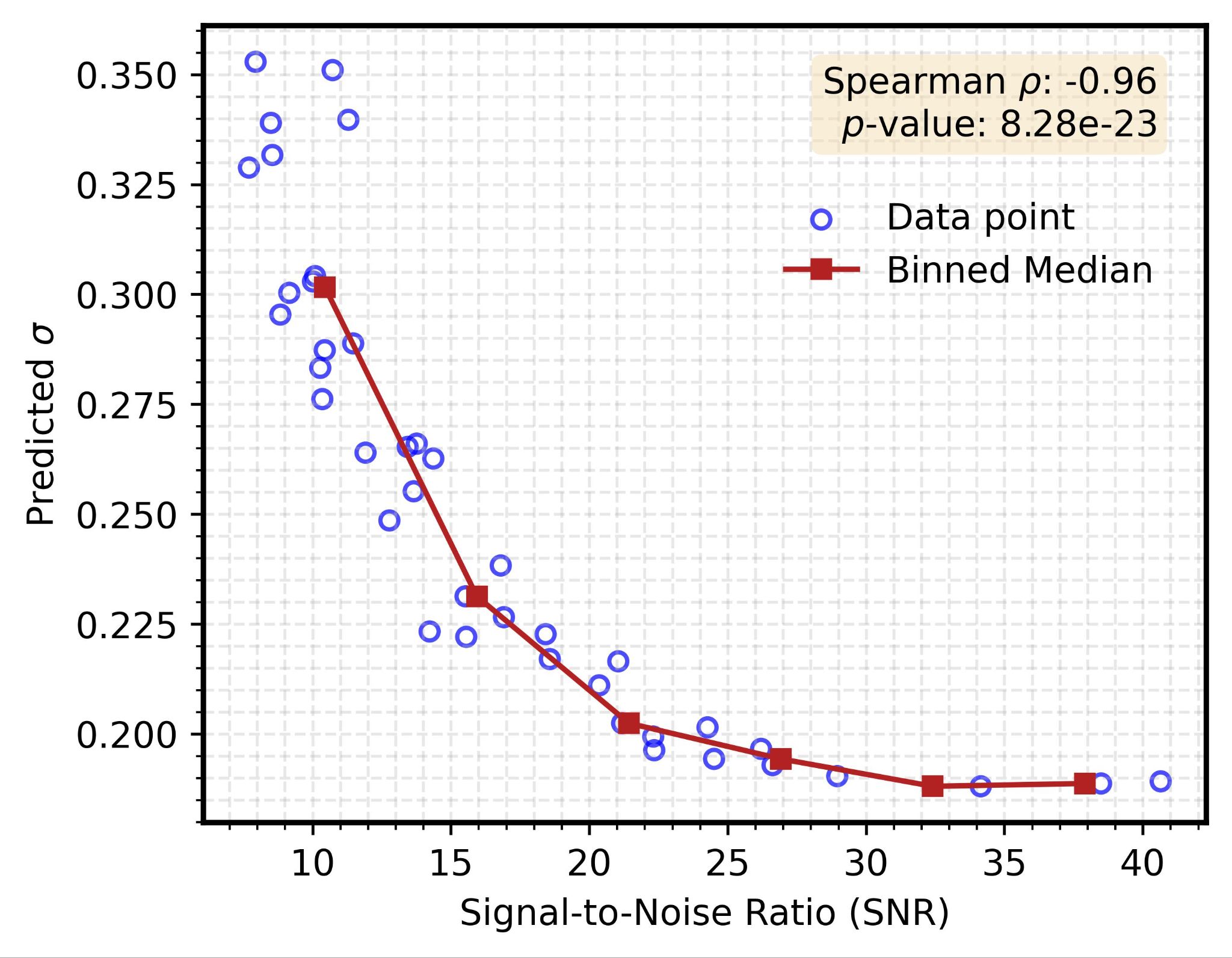}{0.47\textwidth}{}
    }
    \caption{Relationship between predicted uncertainty parameter ($\sigma$) and SNR for two representative cases. Left plot displays the variation of $\sigma$ with SNR for simulated events with width of 1 ms and $\tau$ of 5 ms while the right plot is for simulated events with width of 1 ms and $\tau$ of 10 ms.}
    \label{fig:snr_vs_sigma}
\end{figure}

Figure \ref{fig:snr_vs_sigma} shows the variation of uncertainty parameter $\sigma$ with SNR for two representative cases of scattering with scattering timescales of  5 ms (left) and 10 ms (right). As the scatter plots suggest non-linear relationship between $\sigma$ and SNR, we implemented Spearman correlation method to quantify the correlation in terms of correlation coefficient $\rho$. We found very strong negative correlations between $\sigma$ and SNR for both cases with statistical significance test yielding $p$-values of $p=9.8\times10^{-18}$ for the case with $\tau=$5 ms and $p=8.28 \times10^{-23}$ for the case with $\tau=$10 ms. This confirms that observed correlations are highly unlikely by chance. This statistically significant correlation result confirms that MT-GMDN produces heteroscedastic uncertainty estimates  consistent with observational noise properties.

\section{Discussions}
\label{sec:discussion}
This work demonstrates that FRB morphology can be directly used by a deep learning framework to efficiently and precisely estimate astrophysical parameters such as $\tau$ without relying on explicit pulse-shape fitting. The proposed MT-GMDN framework models the full conditional distribution $p(\tau | \textbf{x})$ capturing both resolved scattering and heteroskedastic uncertainty in a statistically principled manner.  

One of the key results is that the model predicts significant fraction of false positives on the test data set from CHIME/FRB \cattwo as seen in Figure \ref{fig:confusion matrix}. This property of the model should not be interpreted as purely an error because the catalog reported $\tau=0$ values are only the lower limits but not physically exact zeros. Figure \ref{fig:fitburst vs mtgmdn} and Appendix \ref{sec:comparison of fitburst and mtgmdn plots} clearly demonstrate that for low SNR and very subtle scattering events the traditional scattering measurement methods might fail. However, MT-GMDN appears to be more sensitive to subtle temporal asymmetries due to low level scattering signatures which are not captured by traditional fitting approaches when operating near the instrumental resolution limit or low SNRs.

The choice of conditional median as the primary point estimate of $\tau$ instead of mean has been motivated by various factors. One of the main reasons being its robustness to a heavy-tailed lognormal distribution. While the mean is strongly influenced by the large scattering tails, the median often faithfully represents typical scattering timescale imposed by the data. Also, the median of the lognormal corresponds to the exponentiated mean of the underlying normal distribution which allows a convenient construction of the confidence intervals on the point estimation of $\tau$.

\section{Limitations and Future Works}
\label{sec:limitation and future work}
Despite our MT-GMDN yielding promising results, there are still several limitations that motivates future work. The first limitation is that the model was trained on a very small size of training data set and there were modest number of samples with high $\tau$ values ($>100$ ms). This may hinder the model's performance on accurately characterizing highly scattered events. Furthermore,  the model was trained exclusively on total intensity data from CHIME/FRB Catalog 2, generalization to high resolution CHIME/FRB data (baseband data) and also cross instrumental generalization remains to be investigated. Another notable constraint is the model's maximum temporal window (context length) of 162 time samples. This limits the characterization of events requiring larger temporal extent.

To enhance the model performance, future work will involve training the model by augmenting the CHIME/FRB catalog dataset with synthetic bursts injected into real telescope noise. We also plan to incorporate high resolution data sets in the training process allowing the model to better capture subtle burst structures. Additionally, we plan to extend the model architecture to estimate other morphological characteristics beyond the current implementation. 

\section{Conclusion}
\label{sec:conclusion}

In this work, we have presented a Multimodal Transformer Based Generic Mixture Density Network to infer an astrophysical parameter $\tau$ from burst morphology encoded in a dynamic spectrum. The model jointly estimates the probability of unresolved scattering and the conditional distribution of $\tau$ and thus yielding robust point estimates alongside well-calibrated uncertainty bounds.

The introduction of multimodal framework that leverages complementary information from both temporal domain and spectral domain  solves the fundamental challenge of limited dataset to train the transformer based architecture. Our probabilistic mixture density formulation provides a robust solution to handle zero-inflated data set during training thus allowing accurate modeling of both scattered  and unresolved scattered signals. The successful application of the model on real CHIME/FRB test data set and the synthetic test data set demonstrates the robustness and scientific utility of our framework.

The proposed framework offers a non-parametric alternative to classical scattering pipelines and can be implemented to automate the large-scale scattering inference. The classical methods are still useful for verification purposes. 

\section*{acknowledgments}
We acknowledge that CHIME is located on the traditional, ancestral, and unceded territory of the Syilx/Okanagan people.

We are grateful to the staff of the Dominion Radio Astrophysical Observatory, which is operated by the National Research Council of Canada.  CHIME is funded by a grant from the Canada Foundation for Innovation (CFI) 2012 Leading Edge Fund (Project 31170) and by contributions from the provinces of British Columbia, Qu\'ebec and Ontario. The CHIME/FRB project is funded by a grant from the CFI 2015 Innovation Fund (Project 33213) and by contributions from the provinces of British Columbia and Qu\'ebec, and by the Dunlap Institute for Astronomy and Astrophysics at the University of Toronto. Additional support is provided by the Canadian Institute for Advanced Research (CIFAR), McGill University and the McGill Space Institute thanks to the Trottier Family Foundation, and the University of British Columbia. The CHIME/Pulsar instrument hardware is funded by the Natural Sciences and Engineering Research Council (NSERC) Research Tools and Instruments (RTI-1) grant EQPEQ 458893-2014.

 M.N. is a Fonds de Recherche du Quebec - Nature et Technologies (FRQNT) postdoctoral fellow. M.W.S acknowledges support from the French government under the France 2030 investment plan, as part of the Initiative d’Excellence d’Aix-Marseille Université - AMIDEX (AMX-23-CEI- 088). P.S. acknowledges the support of an NSERC Discovery Grant (RGPIN-2024-06266).

\software{
\texttt{Numpy} \citep{harris2020array},
\texttt{Matplotlib} \citep{Hunter:2007},
\texttt{PyTorch} \citep{paszke2019pytorch},
\texttt{torchvision} \citep{torchvision2016},
\texttt{pandas} \citep{reback2020pandas},
\texttt{seaborn} \citep{Waskom2021},
\texttt{fitburst}  \citep{Fonseca2024}, 
\texttt{simpulse} \citep{smith2026simpulse}
}

\section*{Data and Software Availability}

The source code used in this work is publicly available on GitHub\footnote{
\url{https://github.com/kharelb/Scattering-Timescale-Estimation-With-Transformers}
}.

\appendix
\section{Breakdown of model architecture} \label{sec: breakdowm of model architecture}
\subsection{Dynamic Spectrum Transformer} \label{sec:dynamic spectrum transformer}
The dynamic spectrum transformer ingests dynamic spectrum which is tokenized along $n$-dimensional time axis with each token being $d$-dimensional frequency vector:
$$
\textbf{x}_{\textrm{DS}} \in \mathbb{R}^{n \times d}.
$$
\begin{figure}[h!]
    \centering    \includegraphics[width=0.9\linewidth]{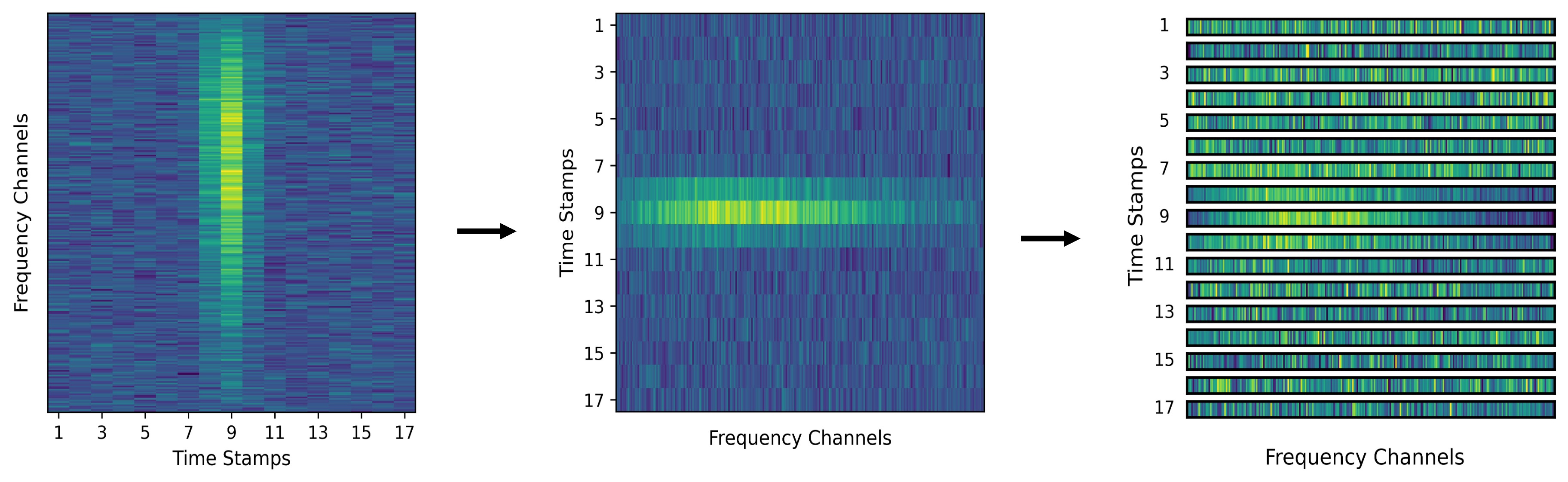}
    \caption{Illustration of the tokenization scheme for a dynamic spectrum. Each time stamp acts as a distinct token and the frequency intensities vector at that time stamp serves as token embedding referred to as d-dimensional frequency vector.}
    \label{fig:ds token embedding}
\end{figure}
This choice of dynamic spectrum input preserves the frequency details for each time stamp and allows the self attention mechanism to model the long range temporal dependencies along with frequency dependent temporal smearing. 

The dynamic spectrum transformer is similar to the standard transformer encoder \citep{vaswani2023attentionneed}  consisting of two encoder layers with each encoder layer containing 4 self-attention heads which performs full self-attention. We experimented with higher number of heads (8, 16) but there was no significant change in the performance of the model yielding the similar values of $R^2$ as given by 4 heads.

\subsection{Timeseries Transformer} \label{sec: timeseries transformer}
\begin{figure}[t!]
    \centering  \includegraphics[width=0.75\linewidth]{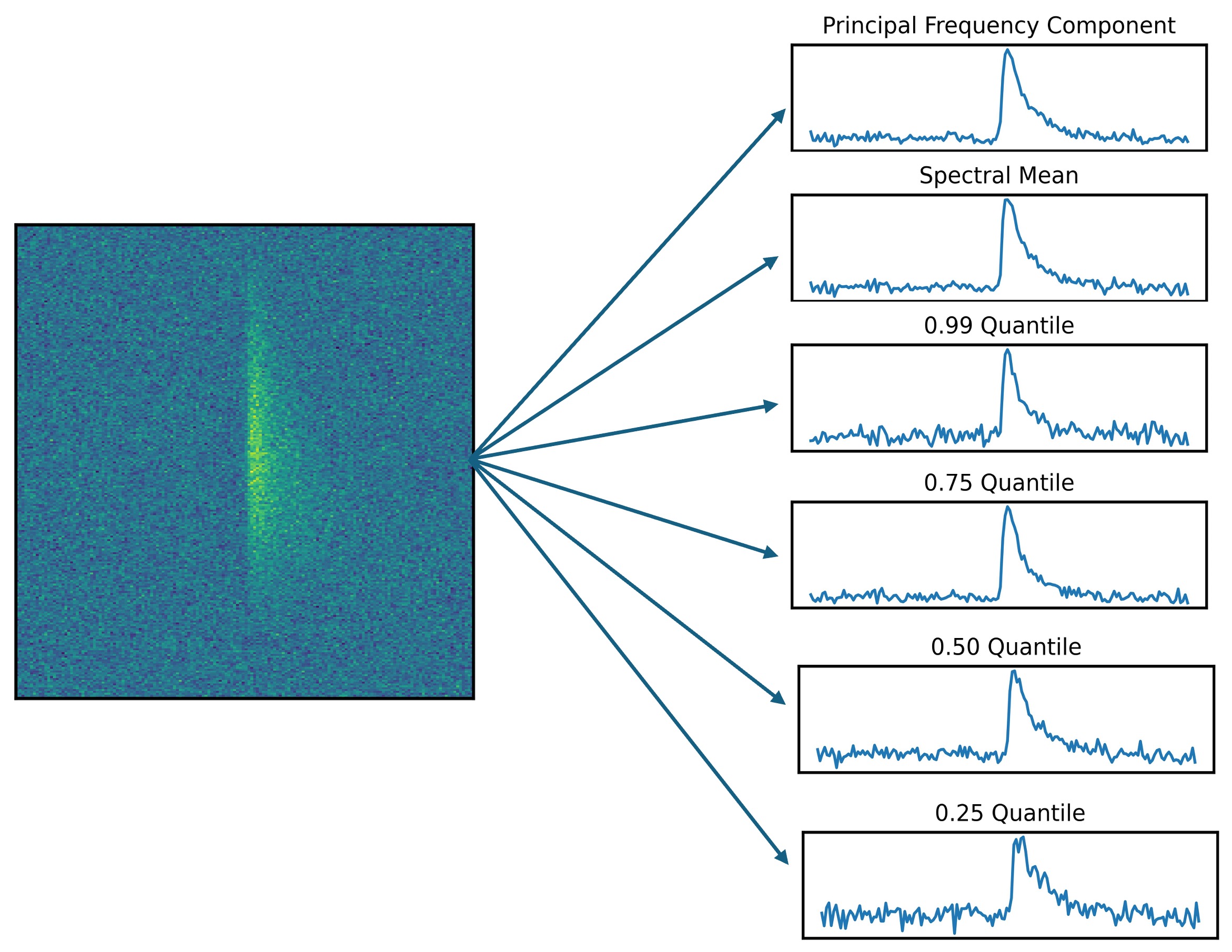}
    \caption{Timeseries (pulse profile) created by different dimensionality reduction techniques along the frequency dimension.}
    \label{fig:timeseries approaches}
\end{figure}

We can compute timeseries from a dynamic spectrum by performing dimensionality reduction along the frequency axis. Some of the timeseries computation methods include simple spectral averaging, projection onto principal frequency components using principal component analysis (PCA), and quantile computation across frequency channels. Timeseries obtained by different approaches as discussed are shown in Figure \ref{fig:timeseries approaches}. Timeseries data thus can be represented as
$$
\textbf{x}_{\textrm{TS}} \in \mathbb{R}^{n}.
$$
The timeseries representation of the dynamic spectrum captures global temporal asymmetry and scattering tails while supressing frequency-dependent structure and thus provides a complementary information to the dynamic spectrum.

\begin{figure}
    \centering
    \includegraphics[width=0.75\linewidth]{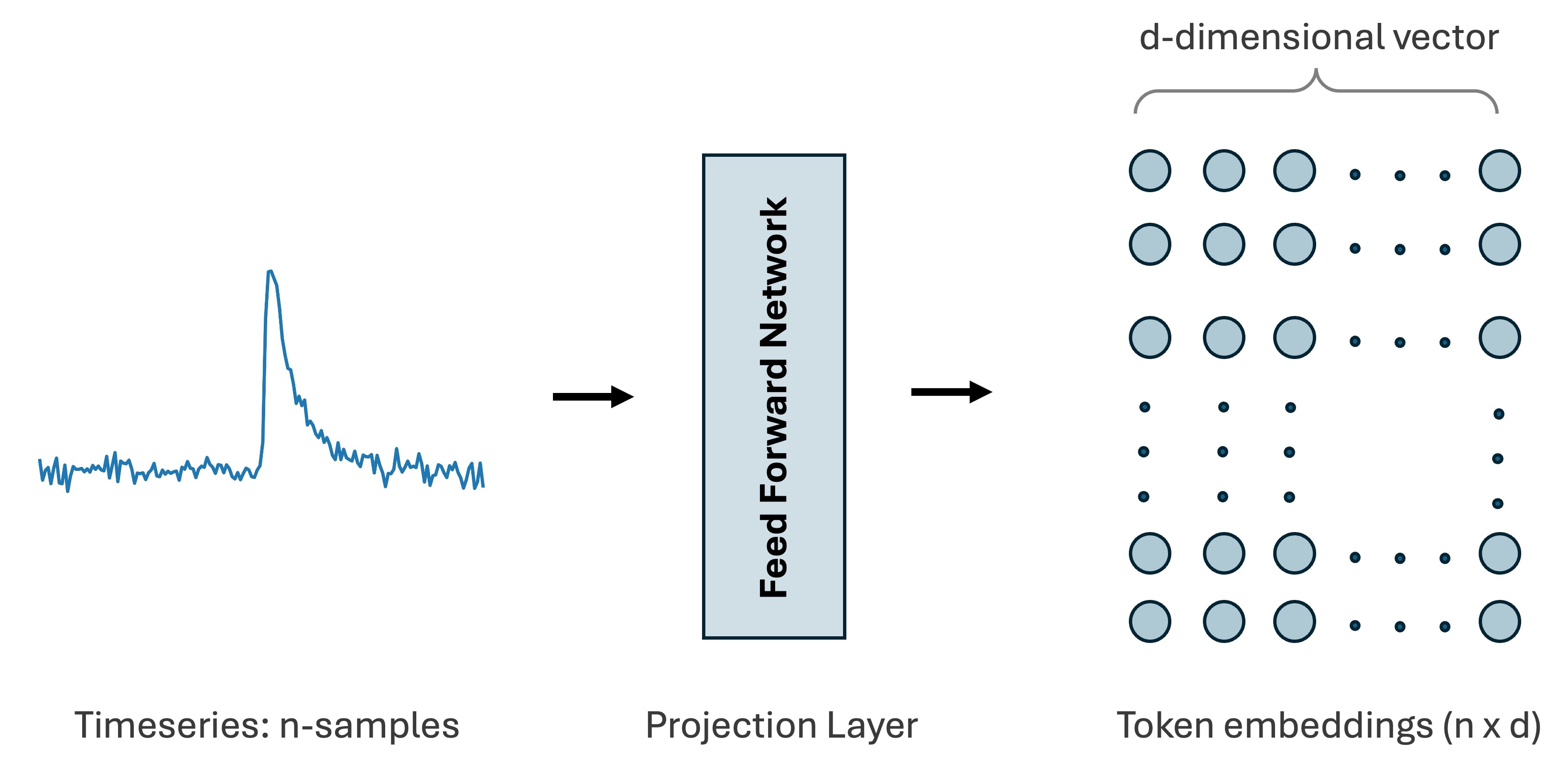}
    \caption{Illustration of the projection layer where each time sample is mapped to $d$-dimensional learnable embedding vector by a simple feed forward neural network.}
    \label{fig:projection layer}
\end{figure}

The projection layer in timeseries transformer projects each scalar timeseries value to $d$-dimensional vector with a simple feed forward network. Each projected vector corresponding to each time sample in timeseries serves as token embedding which is illustrated in Figure \ref{fig:projection layer}.

The timeseries transformer has the same number of encoder layers and attention heads as the dynamic spectrum transformer. This transformer thus focuses on capturing pulse asymmetry, tail decay structure, temporal sub-components and burst duration variations. 

\subsection{Positional encoding} \label{sec:latent representation and fusion}
To make both of the transformers aware of temporal position, positional embedding vectors were added to both dynamic spectrum representation token embeddings and timeseries representation token embeddings. The same positional embeddings were used for both transformers.
\begin{equation*}
\begin{aligned}
\Tilde{\textbf{x}}_{\textrm{DS}} = \textbf{x}_{\textrm{DS}} + \textbf{r}  \\
\Tilde{\textbf{x}}_{\textrm{TS}} = \textbf{x}_{\textrm{TS}} + \textbf{r}
\end{aligned}
\end{equation*}

Upon experimenting, learnable positional embedding yielded better result in our case than sinusoidal positional embedding introduced by \cite{vaswani2023attentionneed}.

\subsection{Latent representation and fusion}
Each transformer generates a sequence of context rich embeddings of the same shape as input:
$$
\textbf{z}_{\textrm{DS}} \in \mathbb{R}^{n \times d}, \hspace{0.5cm} \textbf{z}_{\textrm{TS}} \in \mathbb{R}^{n \times d}.
$$
These two feature rich latent representation outputs from the encoders are fused by concatenation:
\begin{equation*}
    \textbf{z} = [\textbf{z}_{\textrm{DS}}; \textbf{z}_{\textrm{TS}}] \in \mathbb{R}^{n \times 2d}
\end{equation*}
This strategy preserves the modality specific information and also prevents interference between the temporal and spectral features extracted. 

\subsection{Projection and attention pooling}\label{sec:projection and attention pooling}
The fused encoder outputs are then passed through linear projection layer to produce output
\begin{equation}
    \Tilde{\textbf{z}} = \textbf{W}^T\textbf{z} + b,
\end{equation}
where $\Tilde{\textbf{z}} \in \mathbb{R}^{n \times d}$ and $\textbf{W}$ and $\textbf{b}$ are weights and bias associated with the linear transformation. 

We utilize an attention based pooling mechanism to aggregate the output from the last projection layer to a single vector for regression head. We first project each vector from $\Tilde{\textbf{z}}$ corresponding to time step to a scalar score using learnable linear transformation
\begin{equation*}
    \textbf{e} = \textbf{W}^T\Tilde{\textbf{z}} + b,
\end{equation*}
where $\textbf{W}$ and $b$ are learnable parameters. The learned scores are then normalized across the temporal dimension using the softmax function to obtain attention weights $\alpha_t$:
\begin{equation*}
    \alpha_t = \frac{\text{exp}(e_t)}{\sum_{t=1}^n\text{exp}(e_t)}
\end{equation*}

Finally, a weighted sum is performed to produce a global representation
\begin{equation*}
    h = \sum_{t=1}^n\alpha_t\Tilde{\textbf{z}_t}.
\end{equation*}
This approach allows the model to weight different time stamps dynamically based on their relevance.
\subsection{Regression head}
\begin{figure}
    \centering
    \includegraphics[width=0.5\linewidth]{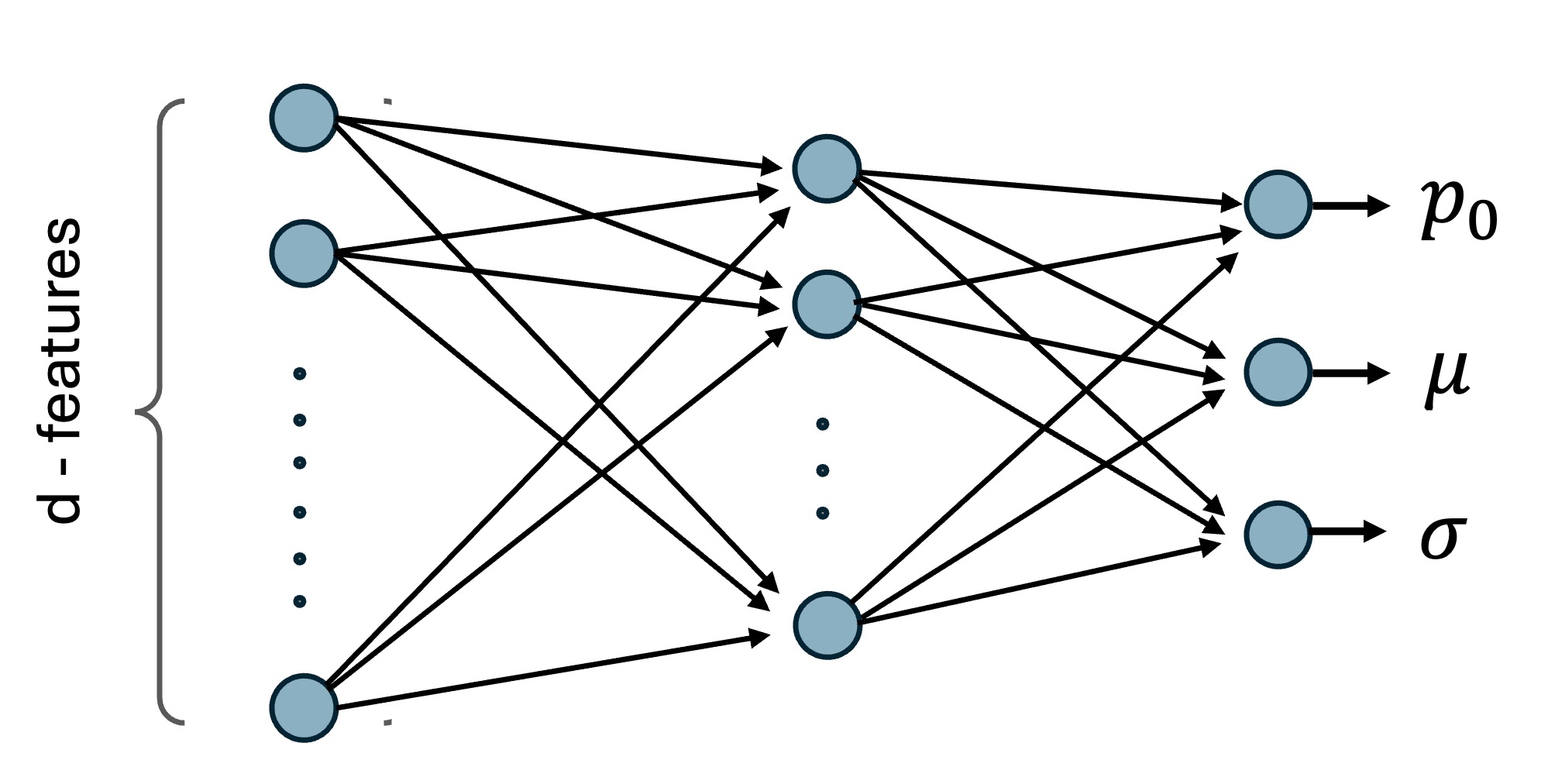}
    \caption{A schematic diagram of regression head. The $d$-dimensional output from the preceding attention pooling layer is passed through a linear layer to obtain the parameters of the mixture density as outputs.}
    \label{fig:regression head}
\end{figure}

The regression head accepts a latent vector $h \in \mathbb{R}^{d}$ from the preceding attention pooling layer. The latent vector is then passed through a linear layer to give an output $\Tilde{h}$. It is then fed into three parallel linear heads as shown in Figure \ref{fig:regression head} which are responsible for distinct distribution parameters of the Bernoulli and lognormal distributions in the mixture density. 
The three outputs from the regression head are further discussed below.
\begin{description}
    \item[Bernoulli head ($f_{p_0}$)] It predicts the probability of zero scattering timescale i.e. the Bernoulli probability of the mixture density. 
    \begin{equation*}
        p_0 = \boldsymbol{\sigma}(f_{p_0}(\Tilde{h}))
    \end{equation*}
    Where $\boldsymbol{\sigma}$ is standard sigmoid function
    \begin{equation*}
        \boldsymbol{\sigma}(x) = \frac{1}{1 + \textrm{exp}(-x)}.
    \end{equation*}

    \item[Location head ($f_{\mu}$)] It predicts the location parameter of the lognormal distribution in the mixture density.
    \begin{equation*}
        \mu = f_{\mu}(\Tilde{h})
    \end{equation*}

    \item[Scale head ($f_{\sigma}$)] It predicts the scale parameter of the lognormal in the mixture density. 
    \begin{equation*}
        \sigma = \textrm{softplus}(f_{\sigma}(\Tilde{h}))
    \end{equation*}
    Where,
    \begin{equation*}
        \textrm{softplus}(x) = \ln(1 + e^x)
    \end{equation*}
\end{description}
\section{Optimization function}\label{sec:optimization function}
The optimization task in our case is maximizing the likelihood with respect to the mixture density given by Equation \ref{eq:GMDN}. So our objective (error) function as seen in Equation \ref{eq:maximize GMDN} is
\begin{equation}
\label{eq:mixture likelihood}
    E(\boldsymbol{\theta}) = \prod_{i=1}^{N}\left\{\sum_{k=1}^K\pi_k(\textbf{x}_i)f_k(\tau_i | \textbf{x}_i;g_{k}(\textbf{x}_i))\right\}.
\end{equation}
Where, $\boldsymbol{\theta}$ are the weights of the neural network. Using Equations \ref{eq:mixture components}, Equation \ref{eq:mixture likelihood} becomes
\begin{equation}
    E(\boldsymbol{\theta}) = \prod_{i=1}^{N}\left\{p_0(\textbf{x}_i;g_{\boldsymbol{\theta}}(\textbf{x}_i))\delta(\tau_i) + (1-p_0(\textbf{x}_i; g_{\boldsymbol{\theta}}(\textbf{x}_i))f_{\textrm{LN}}(\tau_i |\textbf{x}_i;g_{\boldsymbol{\theta}}(\textbf{x}_i) \right\}.
\end{equation}
If $p(\tau_i | \textbf{x}_i; g_{\boldsymbol{\theta}}(\textbf{x}_i))$ is the probability of observing scattering timescale $\tau_i$ for a given sample $\textbf{x}_i$ then there arises two cases:
\begin{description}
    \item[Case I: $\tau_i = 0$]
    \begin{equation*}
    p(\tau_i | \textbf{x}_i; g_{\boldsymbol{\theta}}(\textbf{x}_i)) = p_0(\textbf{x}_i;g_{\boldsymbol{\theta}}(\textbf{x}_i))
    \end{equation*}
    \item[Case II: $\tau_i>0$] \
    \begin{equation*}
         p(\tau_i | \textbf{x}_i; g_{\boldsymbol{\theta}}(\textbf{x}_i)) = (1-p_0(\textbf{x}_i; g_{\boldsymbol{\theta}}(\textbf{x}_i))f_{\textrm{LN}}(\tau_i |\textbf{x}_i;g_{\boldsymbol{\theta}}(\textbf{x}_i)
    \end{equation*}
\end{description}
Thus the objective function is:
\begin{equation}
    E(\boldsymbol{\theta}) = \prod_{i=1}^{N} p(\tau_i | \textbf{x}_i; g_{\boldsymbol{\theta}}(\textbf{x}_i))
\end{equation}
Instead of maximizing the above expression directly, it becomes computationally efficient to maximize the log transformed value of the above expression. Since, the logarithm is a monotonically increasing function, maximizing above expression is equivalent to maximizing the log transformed value. Thus our objective function can be written as
\begin{equation}
     E(\boldsymbol{\theta}) = \sum_{i=1}^{N} \ln \left\{ p(\tau_i | \textbf{x}_i; g_{\boldsymbol{\theta}}(\textbf{x}_i))\right\}.
\end{equation}
Maximizing above equation is equivalent to minimizing the negative of it, so the objective function becomes:
\begin{equation}
\label{eq:neg log likelihood}
     E(\boldsymbol{\theta}) = -\sum_{i=1}^{N} \ln \left\{ p(\tau_i | \textbf{x}_i; g_{\boldsymbol{\theta}}(\textbf{x}_i))\right\}.
\end{equation}
Now, let us derive the logarithm of probabilities for both cases $\tau_i=0$ and $\tau_i > 0$
\begin{description}
    \item[Case I: $\tau_i = 0$ ] \
    If probability under the model:
    $$
    p_0(\tau_i = 0 | \textbf{x}_i;g_{\boldsymbol{\theta}}(\textbf{x}_i)) = p_{0, i}
    $$
    Then,
    \begin{equation}
    \ln\{p_0(\textbf{x}_i;g_{\boldsymbol{\theta}}(\textbf{x}_i))\} = \ln p_{0, i}
    \end{equation}

    \item[Case II: $\tau_i>0$]
    In this case:
    \begin{align*}
         &p_0(\tau_i > 0 | \textbf{x}_i;g_{\boldsymbol{\theta}}(\textbf{x}_i)) = (1-p_{0, i})f_{\textrm{LN}}(\tau_i |\textbf{x}_i;g_{\boldsymbol{\theta}}(\textbf{x}_i)) \\
        \implies &\ln p_0(\tau_i > 0 | \textbf{x}_i;g_{\boldsymbol{\theta}}(\textbf{x}_i)) = \ln (1 - p_{0, i}) + \ln f_{LN}(\tau_i | \textbf{x}_i; g_{\boldsymbol{\theta}}(\textbf{x}_i))
    \end{align*}

But since,
\begin{align*}
    &f_{LN}(\tau_i | \textbf{x}_i; g_{\boldsymbol{\theta}}(\textbf{x}_i)) = \frac{1}{\tau_i \sigma \sqrt{2\pi}} \exp\left( - \frac{(\ln \tau_i - \mu_i)^2}{2\sigma_i^2} \right) \\
    \implies &\ln f_{LN}(\tau_i | \textbf{x}_i; g_{\boldsymbol{\theta}}(\textbf{x}_i)) = - \ln(\tau_i) - \ln(\sigma_i) - \frac{1}{2}\ln(2\pi) - \frac{(\ln \tau_i - \mu_i)^2}{2\sigma_i^2}
\end{align*}
So, 
\begin{equation}
    \ln p_0(\tau_i > 0 | \textbf{x}_i;g_{\boldsymbol{\theta}}(\textbf{x}_i)) =  \ln (1 - p_{0, i}) - \ln(\tau_i) - \ln(\sigma_i) - \frac{1}{2}\ln(2\pi) - \frac{(\ln \tau_i - \mu_i)^2}{2\sigma_i^2}
\end{equation}

Finally, our error function given by Equation \ref{eq:neg log likelihood} normalized for data size becomes
\begin{equation}
\label{eq:final neg likelihood}
    \boxed{
     E(\boldsymbol{\theta}) = -\frac{1}{N}\sum_{i=1}^{N} \ln \left\{ p(\tau_i | \textbf{x}_i; g_{\boldsymbol{\theta}}(\textbf{x}_i))\right\}
    }
\end{equation}
With,
\begin{equation}
\label{eq:final likelihood cases}
 \ln \left\{ p(\tau_i | \textbf{x}_i; g_{\boldsymbol{\theta}}(\textbf{x}_i))\right\} =
    \begin{cases}
     \ln p_{0, i}, & \tau_i = 0,\\
      \ln (1 - p_{0, i}) - \ln(\tau_i) - \ln(\sigma_i) - \frac{1}{2}\ln(2\pi) - \frac{(\ln \tau_i - \mu_i)^2}{2\sigma_i^2}, & \tau_i > 0
    \end{cases}
\end{equation}
\end{description}
Thus the Equation                        \ref{eq:final neg likelihood} is our final objective function which is minimized during training by gradient descent method.

\section{Implementation details and hyperparameters}
\label{sec:hyperparameters}
In this section, we provide the specific configuration and hyperparameter settings used during training to ensure the reproducibility of our results. The model architecture was implemented with deep learning framework \texttt{PyTorch} \citep{paszke2019pytorch} and was trained on \texttt{NVIDIA A100 Tensor Core} GPU provided by \texttt{Dolly Sods}\footnote{\href{https://docs.hpc.wvu.edu/text/84.DollySods.html}{A high performance computing cluster at West Virginia University.}}. All the model architectural parameters were set through manual tuning on the validation dataset. The checkpoint that achieved the minimum validation loss was the best model with robust configuration for generalization presented in Table \ref{tab:hyperparams}.

\begin{deluxetable*}{llcl}
\tablecaption{MT-GMDN Hyperparameter Configuration \label{tab:hyperparams}}
\tablewidth{0pt}
\tablehead{
\colhead{Category} & \colhead{Hyperparameter} & \colhead{Value} & \colhead{Description}
}
\startdata
Architecture & Transformer Heads & 4 & Multi-head attention count \\
             & Encoder Layers & 3 & Number of stacked transformer blocks \\
             & Embedding Dim ($d_{\rm model}$) & 256 & Latent vector size \\
             & FFNN Dimension & 2048 & Feed-forward hidden layer size \\
             & Dropout Rate & 0.1 & Regularization to prevent overfitting \\
             & MDN Hidden Units & $[256, 256]$ & Neurons in the mixture density head \\
\hline
Training     & Batch Size & 32 & Number of samples per iteration \\
             & Learning Rate & $1 \times 10^{-4}$ & Learning rate for optimizer \\
             & Weight Decay & $0.09$ & $L_2$ regularization penalty \\
             & Max Epochs & 400 & Total training iterations \\
             &Optimizer & AdamW & Optimization algorithm \\
\hline
Data         & Temporal Window & 162 & Maximum input time samples \\
             & Frequency Channels & 256 & Spectral channels of dynamic spectra \\
\enddata
\tablecomments{The hyperparameters for both dynamic spectrum and timeseries transformers are same.}
\end{deluxetable*}

\section{Model performance with different timeseries approaches}
\label{sec:model performance on different timeseries}
The estimation of scattering timescale with MT-GMDN relies on reduction of two dimensional dynamic spectrum into a corresponding one dimensional timeseries. While there are different statistical approaches to generate timeseries from a dynamic spectrum, in this work we utilized spectral mean, frequency quantiles, and first principal frequency component using principal component analysis (PCA) and evaluated their performance after training. The best performing model was achieved on spectral mean approach and is explained in Section \ref{sec:training and results}.
\begin{figure*}[ht!]
\gridline{
    \fig{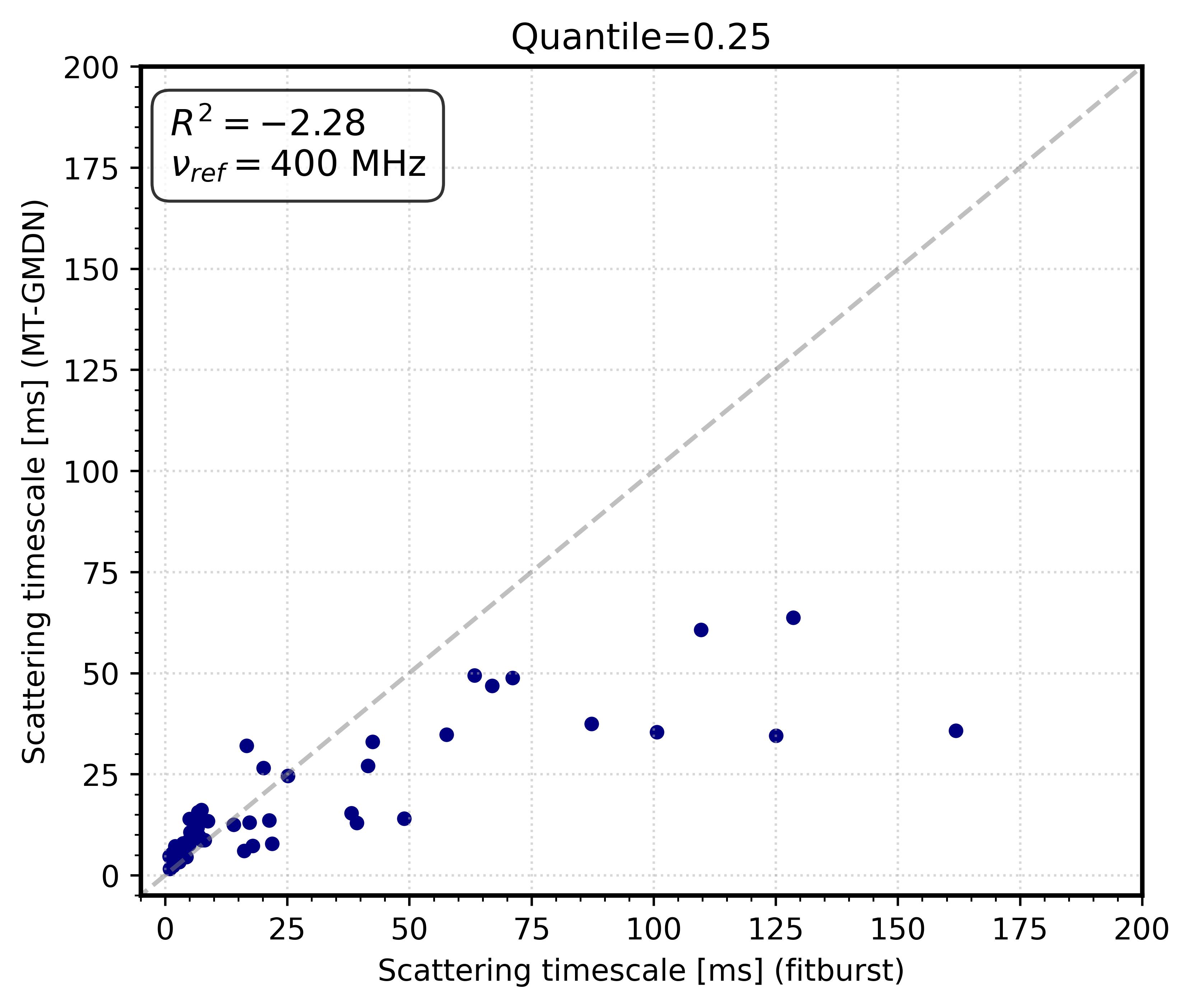}{0.48\textwidth}{}
    \fig{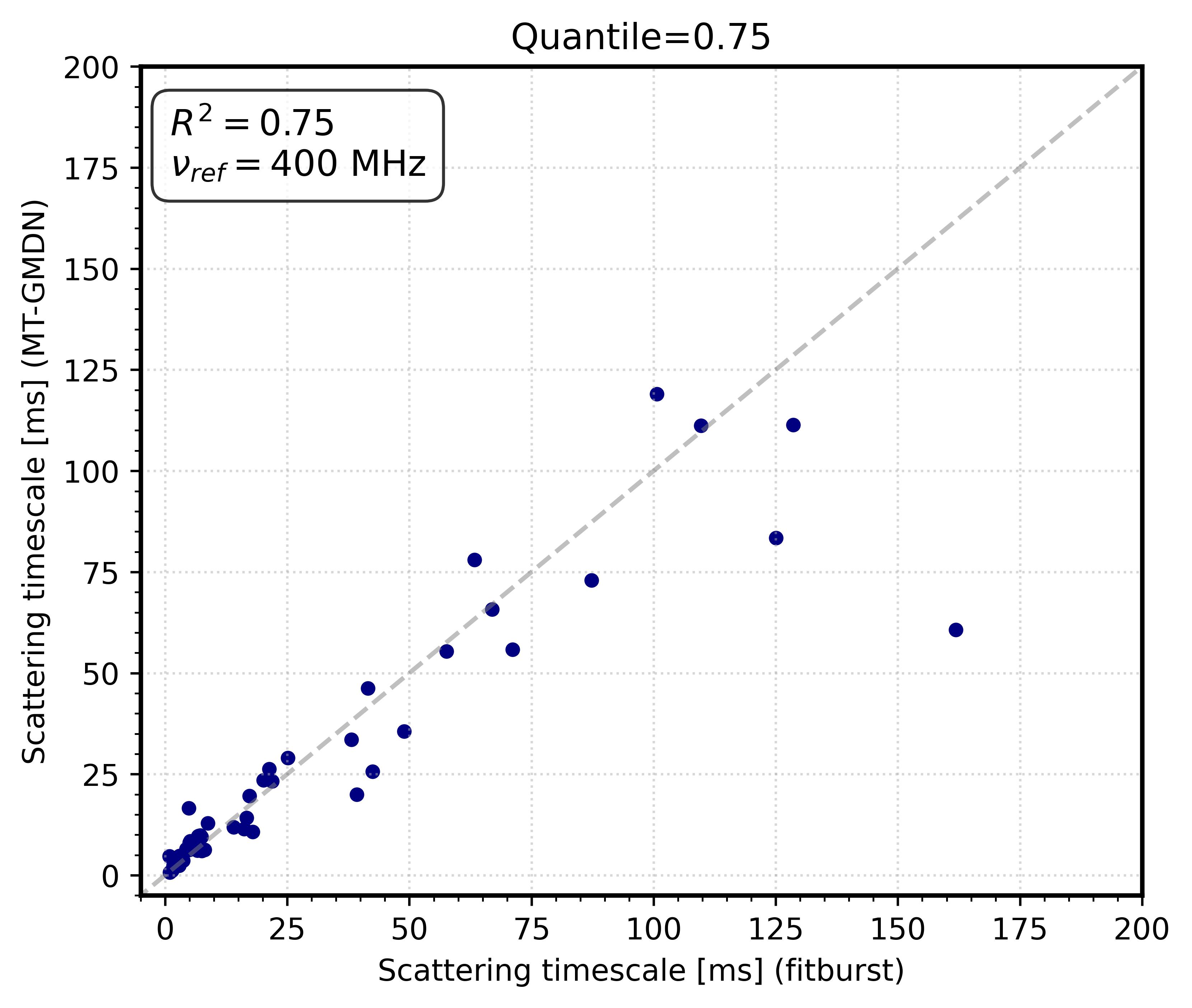}{0.48\textwidth}{}
}
\gridline{
    \fig{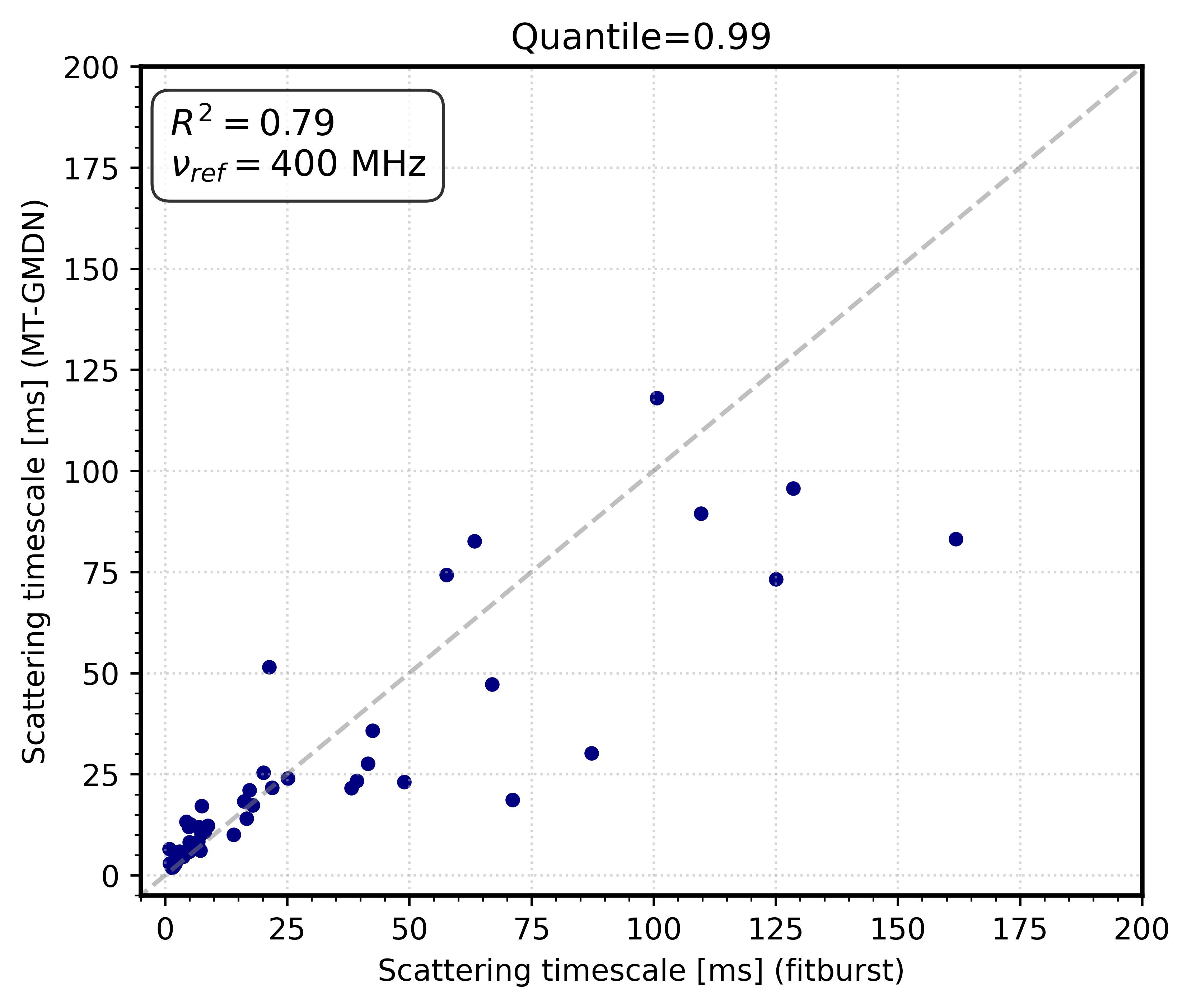}{0.48\textwidth}{}
    \fig{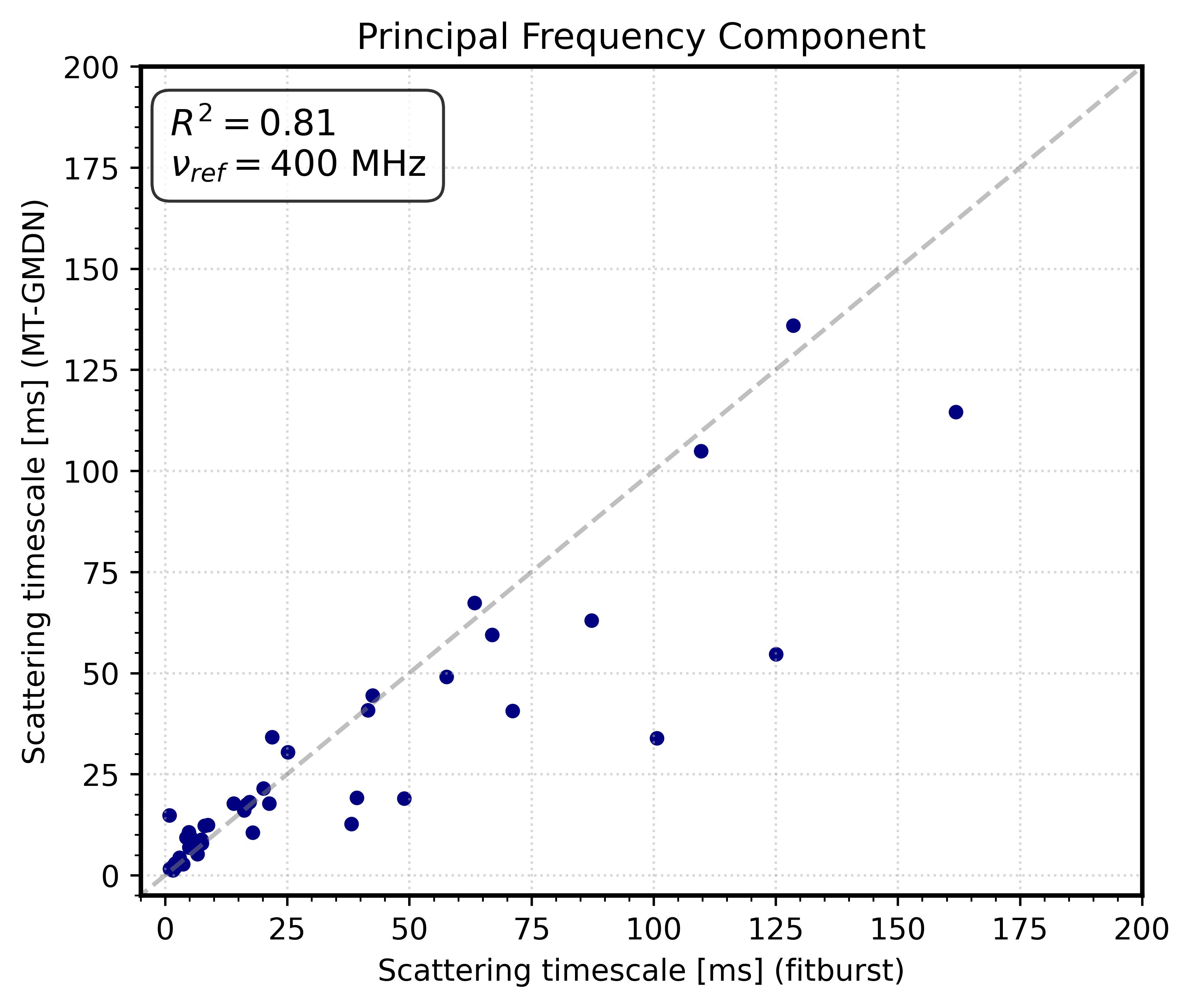}{0.48\textwidth}{}
}
\caption{Evaluation of MT-GMDN performance across four different timeseries extraction methods.}
\label{fig:timeseries comparison}
\end{figure*}

Figure \ref{fig:timeseries comparison} presents the evaluation of different alternative approaches other than spectral mean on the real test data set. The results suggests that these methods peform reliably at low scattering timescales but they struggle for pulses exhibiting significant scattering tails.

\section{Model Comparison with some individual synthetic bursts}
\label{sec:comparison of fitburst and mtgmdn plots}
\begin{figure*}[ht!]
    \centering
    \includegraphics[width=0.95\textwidth]{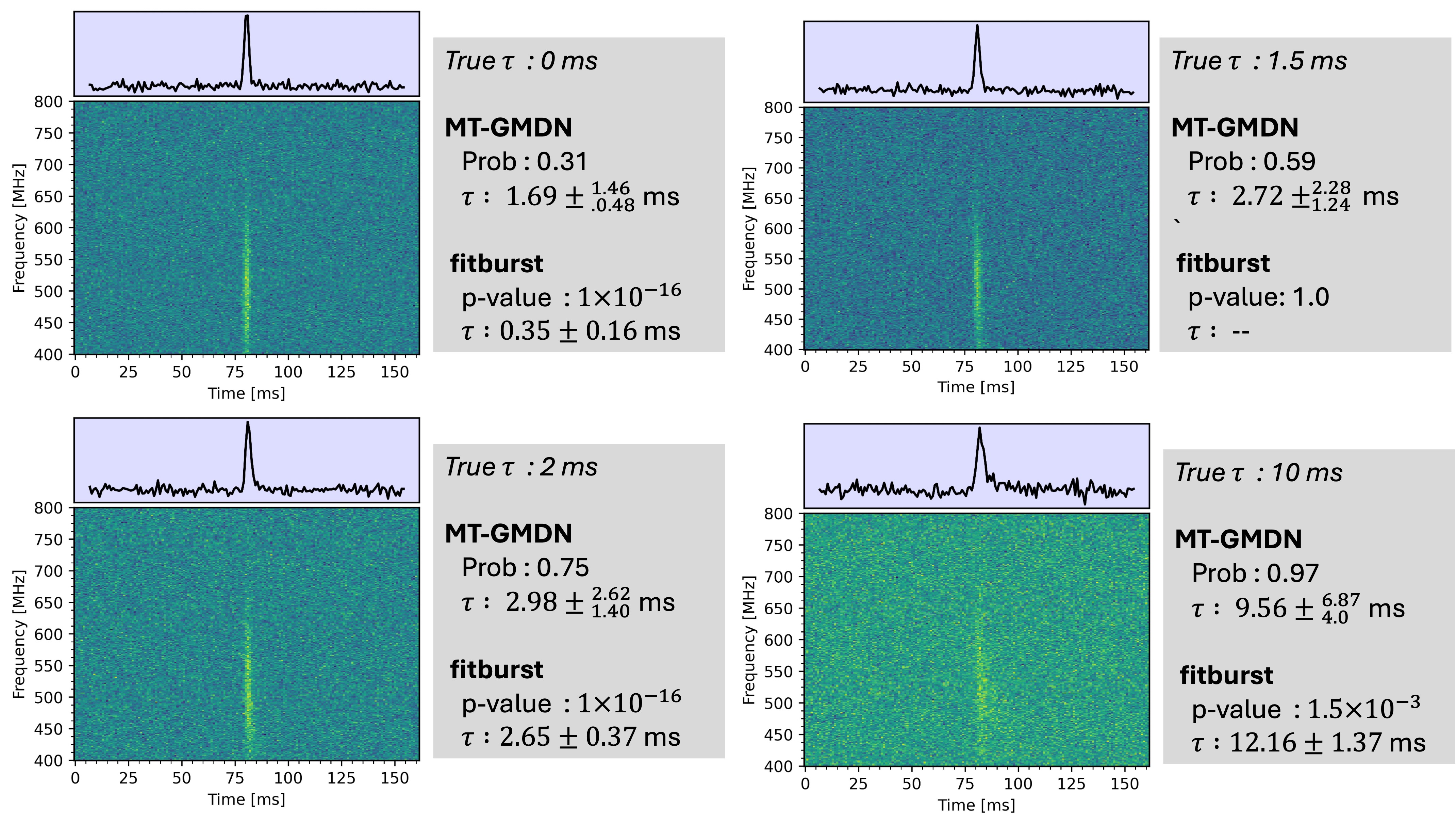}
    
    \vspace{-0.2cm} 
    
    \includegraphics[width=0.95\textwidth]{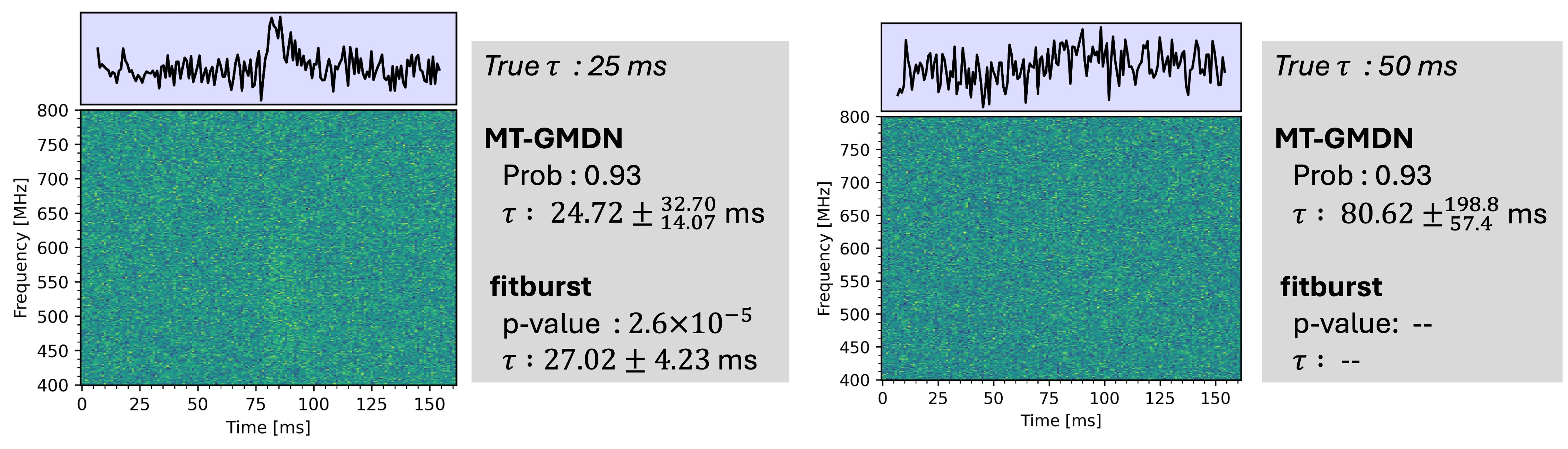}
    
    \caption{Each panel includes two sub-panels with the first sub-panel containing a simulated dynamic spectrum and the corresponding timeseries. While the second sub-panel describes real injected $\tau$ value, the MT-GMDN point estimate with 95\% confidence interval, and the \texttt{fitburst} fit statistics. The uncertainty provided for the \texttt{fitburst} measurement here corresponds to the $1\sigma$ limits. All the $\tau$ values are referenced at 400 MHz.}
    \label{fig:scattering comparison plots}
\end{figure*}

Here, we present a comparison between \texttt{simpulse} generated pulses, \texttt{fitburst} measured values and corresponding p-values of F-test for scattering detections and the estimates from the MT-GMDN. All of the examples in the Figure \ref{fig:scattering comparison plots} have intrinsic width of 1 ms and same fluence values while noise levels vary between the bursts. p-values from \texttt{fitburst} measurements dictates the existence of scattering in a burst while prediction probability of scattering dictates the presence of scattering by MT-GMDN approach. Examining these individual bursts with aforementioned methods provides an insight on how MT-GMDN performs across different burst profiles. We can observe that MT-GMDN maintains prediction accuracy of $\tau$ values in various regimes where \texttt{fitburst} returns non significant p-values (p $>$ 0.05) or even when \texttt{fitburst} fails to detect a burst in a highly scattered low SNR profile.    

\section{Some FRBs with strong noise and radio frequency interference}
\begin{figure}
    \centering
    \includegraphics[width=0.90\linewidth]{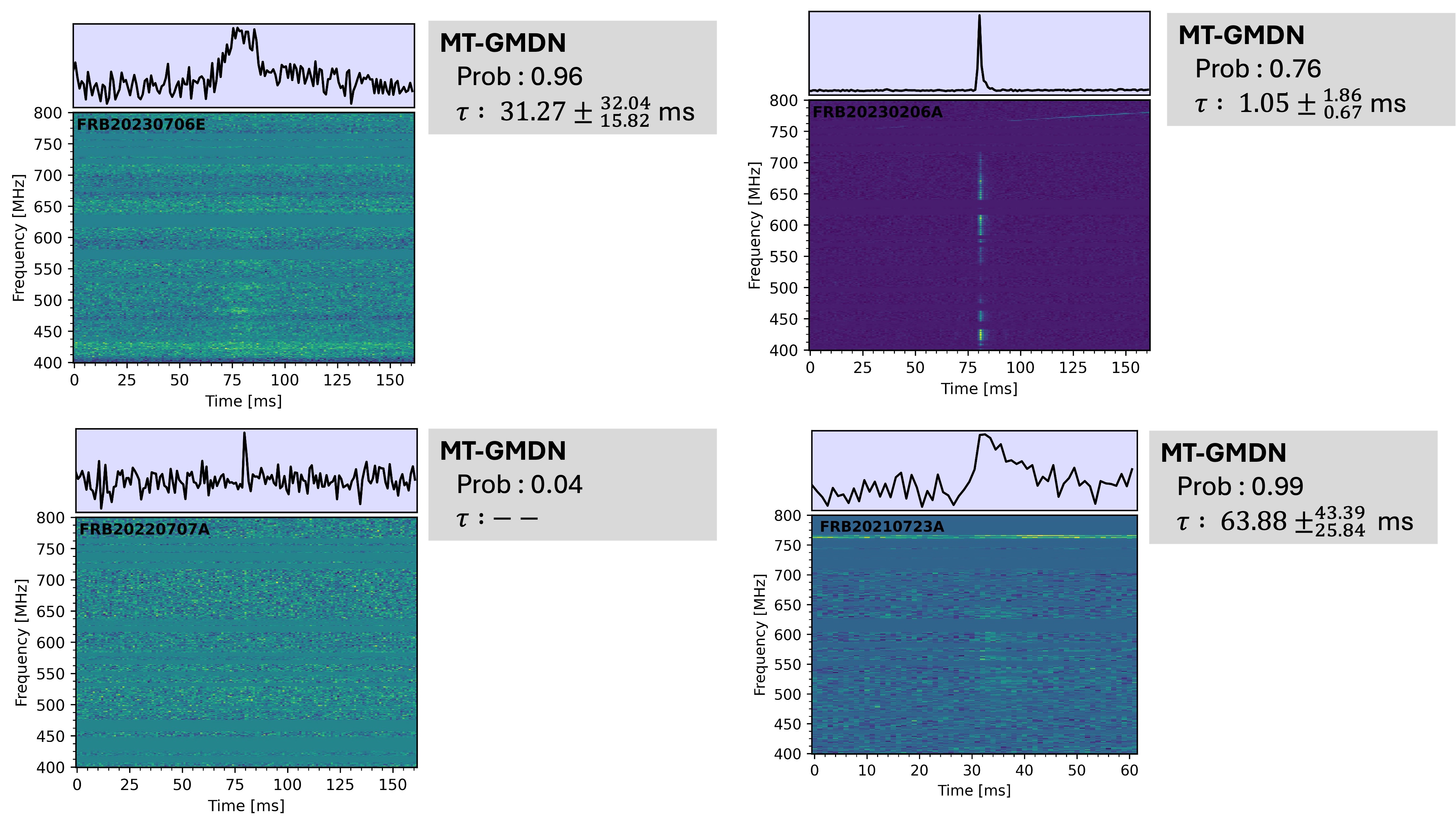}
    \caption{Representative FRBs from the CHIME/FRB Catalog 2 where traditional methods failed to extract physical parameters due to strong noise and RFI. Each panel includes two sub-panels with the first sub-panel (left) containing a dynamic spectrum and the corresponding timeseries, while the second sub-panel (right) contains MT-GMDN estimates. The estimates include probability of scattering, point estimate for $\tau$ value and its $95\%$ predictive interval.}
    \label{fig:real noisy data cat 2}
\end{figure}

To demonstrate significance of MT-GMDN, we present a set of representative FRBs shown in the Figure \ref{fig:real noisy data cat 2} from the CHIME/FRB Catalog 2 that are affected by noise and radio frequency interference (RFI). Due to the strong noise and heavy RFI, extracting any physical information such as $\tau$ is very challenging by traditional methods discussed before. Despite the challenges, MT-GMDN produces visibly stable predictions with increased uncertainty to lower quality observations as displayed in Figure \ref{fig:real noisy data cat 2}.

\section{Alternative Gamma Mixture Density Formulation}
As seen above from results, the lognormal formulation can produce wide upper confidence bounds for events with higher scattering and/or lower signal to noise. This is because, the variance is computed in the log scale and then transformed back to linear scale. A very small variation of $\sigma$ in log scale can lead to large variation in the linear scale. One alternative to resolve the inflated confidence interval upper bound can be the use of distribution in the mixture density formulation with lighter tail than lognormal. Also the distribution should be strictly positive and operating in linear scale. 

Gamma distribution satisfies all of the properties of alternative distribution discussed above. In this section, we examine the effect of distribution on confidence interval width by replacing the lognormal with the gamma distribution in the mixture density formulation in Equation \ref{eq:GMDN}. Gamma distribution for a random variable $x$ is given by 
\begin{equation}
\label{eq:gamma distribution}
    f(x; \alpha, \beta) = \frac{\beta^\alpha x^{\alpha-1} e^{-\beta x}}{\Gamma(\alpha)},
\end{equation}
where $\alpha$ is the shape parameter, $\beta$ is rate parameter and $\Gamma(\alpha)$ is the Gamma function. Following Section \ref{sec:optimization function}, we can easily compute the error function by replacing $f_2(\textbf{x})$ in Equation \ref{eq:mixture components} by the Gamma distribution. The logarithm term in Equation \ref{eq:final likelihood cases} thus gets modified to
\begin{equation}
\log p(\tau_i\mid \mathbf{x}_i)
=
\begin{cases}
\log p_{0,i}, & \tau_i=0,\\[6pt]
\log(1-p_{0,i}) + \alpha_i \log \beta_i - \log \Gamma(\alpha_i) + (\alpha_i-1)\log \tau_i - \beta_i\tau_i, & \tau_i>0.
\end{cases}
\end{equation}

The neural network was then trained to estimate the parameters of the Gamma distribution and the Bernoulli's parameter in the mixture density formulation. Point estimate was computed by the median of the target distribution with the upper and the lower bounds defined by 0.025 and 0.975 quantiles respectively. 
\begin{figure}
    \centering
    \includegraphics[width=0.75\linewidth]{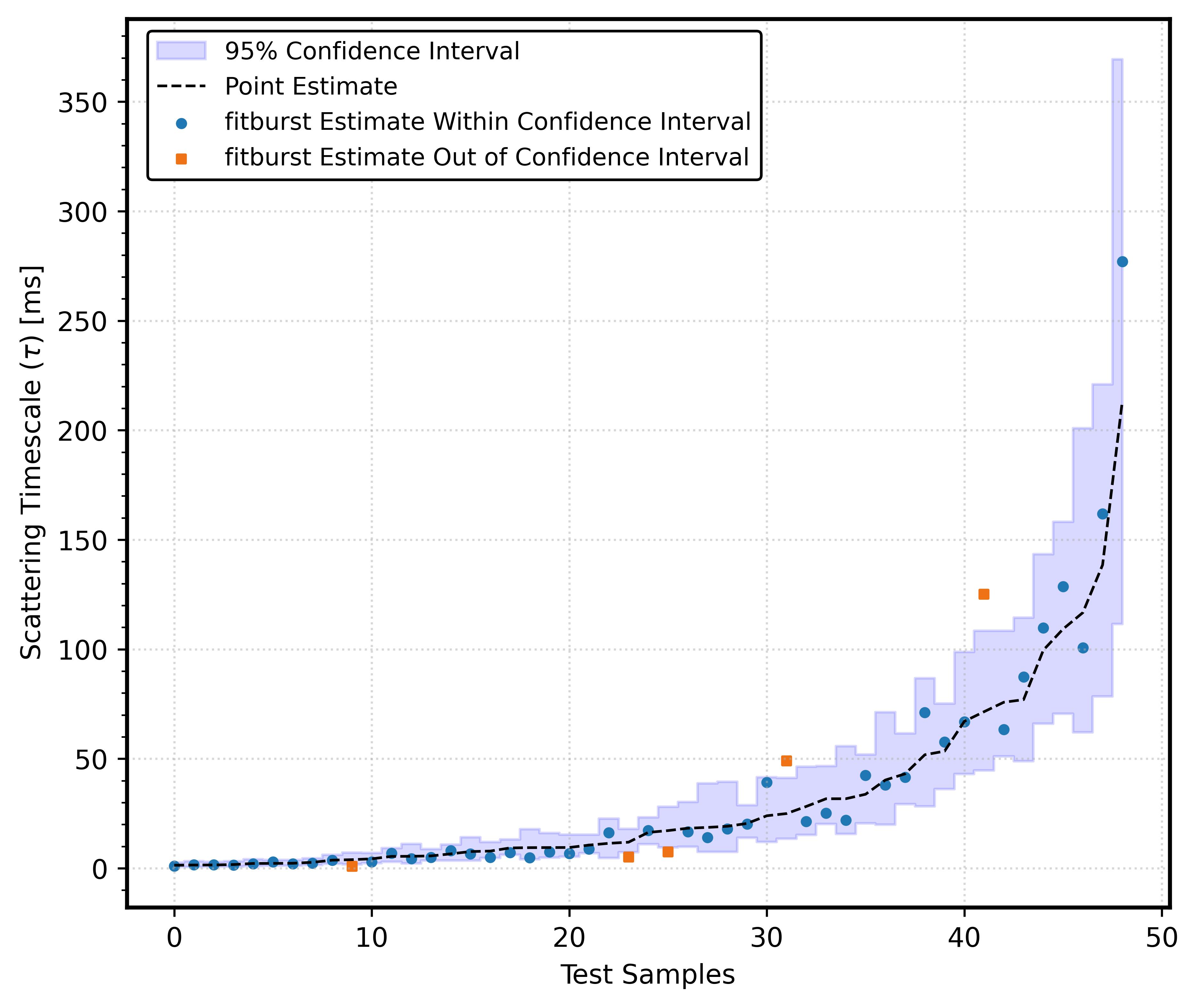}
    \caption{Test set prediction performance with x-axis denoting discrete FRB samples and y-axis scattering timescale at a reference frequency of 400 MHz. The mixture density formulation here implements the Gamma distribution instead of the Lognormal distribution. \texttt{fitburst} measured values are shown as blue dots and MT-GMDN point estimates are represented by black dashed lines while the blue shaded region representing $95\%$ confidence interval. Orange squares denote the FRB samples which are out of the $95\%$ confidence interval.}
    \label{fig:evaluation_ci_mean_gamma}
\end{figure}
Figure \ref{fig:evaluation_ci_mean_gamma} shows per sample comparison of MT-GMDN predictions against the \texttt{fitburst} measured values after replacing the Lognormal distribution by the Gamma distribution in mixture density formulation for the test data set. $R^2$-score of $88\%$ was achieved which represents slight degradation in the point estimation with the Gamma distribution method. However, it can be clearly seen from Figures \ref{fig:evaluation_ci_mean} and \ref{fig:evaluation_ci_mean_gamma} that the width of the confidence interval in Gamma distribution method produces much more narrower confidence intervals than that of Lognormal distribution method. This can be quantified by the MIW defined in the Table \ref{tab: mcmc vs mtgmdn}. The MIW with the Gamma distribution is 32.74  while with the Lognormal distribution is 56.38. The classification performances with both the Gamma and Lognormal distributions in the mixture density formulation remained almost identical since Bernoulli's distribution was used in both of the cases for classification. 

Thus the Gamma component can significantly reduce the inflated confidence intervals for the high scattering and/or low signal to noise dynamic spectrum at the expense of slightly degraded $R^2$-score in the point estimate.              

\bibliography{sample701}{}
\bibliographystyle{aasjournalv7}



\end{document}